\documentclass{sn-jnl}

\usepackage{graphicx}
\usepackage{float}
\usepackage[algo2e,boxruled,vlined,linesnumbered]{algorithm2e}
\usepackage{extraplaceins}
\usepackage{epstopdf}
\usepackage[numbers,sort&compress]{natbib}

\usepackage{amsmath,amssymb}
\usepackage{amsfonts}
\usepackage{mathtools}

\jyear{2021}%

\theoremstyle{thmstyleone}%
\newtheorem{theorem}{Theorem}
\theoremstyle{thmstyletwo}%

\theoremstyle{thmstylethree}%

\begin{document}

\title[The limited-capacity many-to-many point matching in one dimension]{A faster algorithm for the limited-capacity many-to-many point matching in one dimension}


\author[1]{\fnm{Fatemeh} \sur{Rajabi-Alni}}\email{f.rajabialni@aut.ac.ir}

\author[1]{\fnm{Alireza} \sur{Bagheri}}\email{ar\_bagheri@aut.ac.ir}
\author[2]{\fnm{Behrouz} \sur{Minaei-Bidgoli}}\email{b\_minaei@iust.ac.ir}

\affil[1]{\orgdiv{Computer Engineering Department}, \orgname{Amirkabir University of Technology (Tehran Polytechnic)}, \city{Tehran}, \country{Iran}}

\affil[2]{\orgdiv{Department of Computer Engineering}, \orgname{Iran University of Science and Technology}, \city{Tehran}, \country{Iran}}


\abstract{Given two sets $S$ and $T$, a \textit {limited-capacity many-to-many matching} (LCMM) between $S$ and $T$ matches each element $p$ in $S$ (resp. $T$) to at least $1$ and at most $Cap(p)$ elements in $T$ (resp. $S$), where the function $Cap:S\cup T\rightarrow Z>0$ denotes the capacity of $p$. In this paper, we present the first linear time algorithm for finding a minimum-cost \textit{one-dimensional LCMM} (OLCMM) between $S$ and $T$, where $S$ and $T$ are points lying on a line, and the cost of matching $p\in S$ to $q\in T$ equals the $l_2$ distance between $p,q$. Our algorithm improves the previous best-known quadratic time algorithm.}




\maketitle

\section{Introduction}
\label{intro}

Suppose we are given two sets $S$ and $T$, a \textit {many-to-many matching} (MM) between $S$ and $T$ assigns each element of one set to one or more elements of the other set \cite{ColanDamian}. The minimum-cost MM problem has been solved using the Hungarian method in $O(n^3)$ time, where $n$ is the number of elements of $S\cup T$ \cite{Eiter}. Later, an $O(n \log {n})$ time dynamic programming solution was presented for finding a minimum-cost MM between two sets of points on the real line \cite{ColanDamian}. An $O(n^2poly(\log n))$ algorithm is also given for computing a minimum-cost MM between two sets of points in the plane in \cite{Bandyapadhyay}.

The MM has many applications in the real world, including resource allocation in a distributed network such as the internet of things (IoT) and data collection and transmission in wireless sensor networks \cite{Zhang,Liu,Pradip}. Also, finding a comparable control group for a set of treated units in the era of big data and matching elements from two data sets are two important examples of the applications of the MM \cite{Fredrickson,David}.

In practice, in an MM between two sets, each element of one set can not be matched to an infinite number of elements of the other set. A general case of the MM is the \textit{limited capacity} MM (LCMM), where each element has a capacity, i.e., the number of elements that can be matched to each element. Schrijver \cite{Schrijver} proved that a minimum-cost LCMM could be found in strongly polynomial time. A minimum-cost LCMM can be reduced to finding a minimum-weight degree-constrained subgraph in a general graph $H=(V,E)$, and solved in $O(n^2\min(m \log n,n^2))$ time, where $n=\vert V \vert$ and $m=\vert E\vert$ due to \cite{Gabow1983}. A special case of the minimum-cost LCMM problem is the \textit{one-dimensional} LCMM (OLCMM) problem that in which both $S$ and $T$ lie on the real line, and the cost of each matched pair $(p,q)$ with $p\in S$ and $q\in T$ is the Euclidean distance ($l_2$ distance) between $p$ and $q$. In this paper, we give the first linear time algorithm for the minimum-cost OLCMM problem improving the previous best-known $O(n^2)$ time algorithm presented in \cite{Rajabi-Alni}.

 As an example for the OLCMM, consider two sets of sensors and base stations deployed on a line (such as a highway or border). The sensors collect data from the surrounding environment and send the gathered data to the base stations to be processed and analyzed. Each base station can receive data from a limited number of sensors because of its limited processing capacity and the multiple access interference which arises from co-channel sensors \cite{Khalili}. Also, the limited battery charge of each sensor limits the number of base stations to which the sensor can send its collected data. The power consumed by the sensor $p$ for sending its data to the base station $q$ is proportional to the Euclidean distance between $p$ and $q$, i.e. $\|p-q\|$ \cite{Rasti}. We want to send the data collected by sensors to base stations with minimum energy consumption such that each sensor sends its data to at least one base station and each base station receives data from at least one sensor, and also the limited capacities of sensors and base stations are satisfied. Note that our algorithm is a dynamic programming algorithm, which implies that it can run online, where the points of $S$ and $T$ arrive one by one in an online manner \cite{Ahmed}.


\section{Preliminaries}
\label{PreliminSect}
\begin{figure*}
\vspace{0cm}
\hspace{0cm}
 \includegraphics[width=0.9\textwidth]{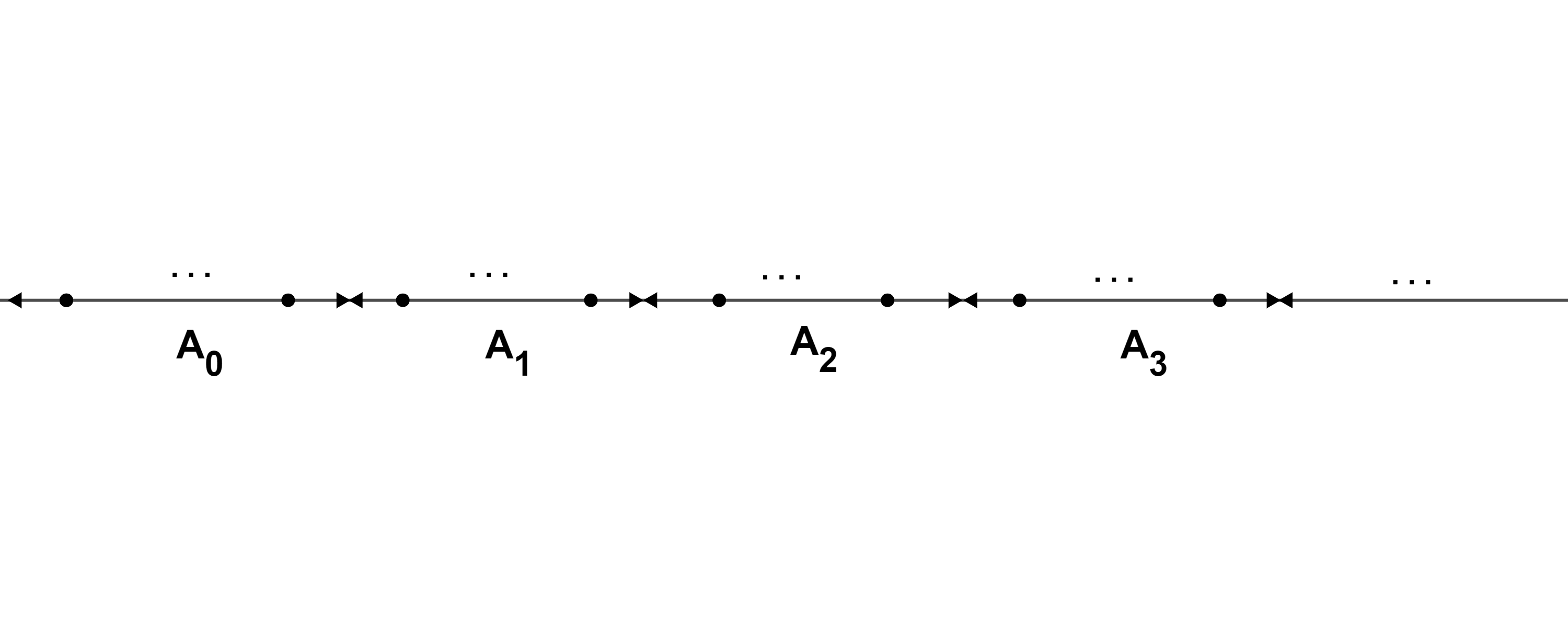}
\vspace{-1.3cm}
\caption{$S \cup T$ is partitioned into maximal subsets $A_0,A_1,A_2,\dots $.}
\label{fig:11}       
\end{figure*}

In this section, we proceed with some useful definitions and assumptions. Let $S=\{s_1, s_2, \dots, s_y\}$ and $T=\{t_1,t_2, \dots, t_z\}$ be two sorted sets of points on the real line such that their elements are in increasing order. Let $S \cup T$ be partitioned into maximal subsets $A_0,A_1,A_2,\dots $ alternating between subsets in $S$ and $T$ such that all the points in $A_i$ are smaller than all the points in $A_{i+1}$ for all $i\geq 0$: the point of the highest coordinate in $A_i$ lies to the left of the point of the lowest coordinate in $A_{i+1}$ (Fig. \ref{fig:11}).
Note that some points may have the same value. In this case, we treat them as different points while partitioning $S\cup T$ into maximal subsets only needs the ordering of points in some directions: we can order equal value points arbitrarily; for instance, order them based on their name. Thus, w.l.o.g. we assume that all points $p \in S\cup T$ are distinct.

\begin{figure*}
\vspace{0cm}
\hspace{0cm}
 \includegraphics[width=1\textwidth]{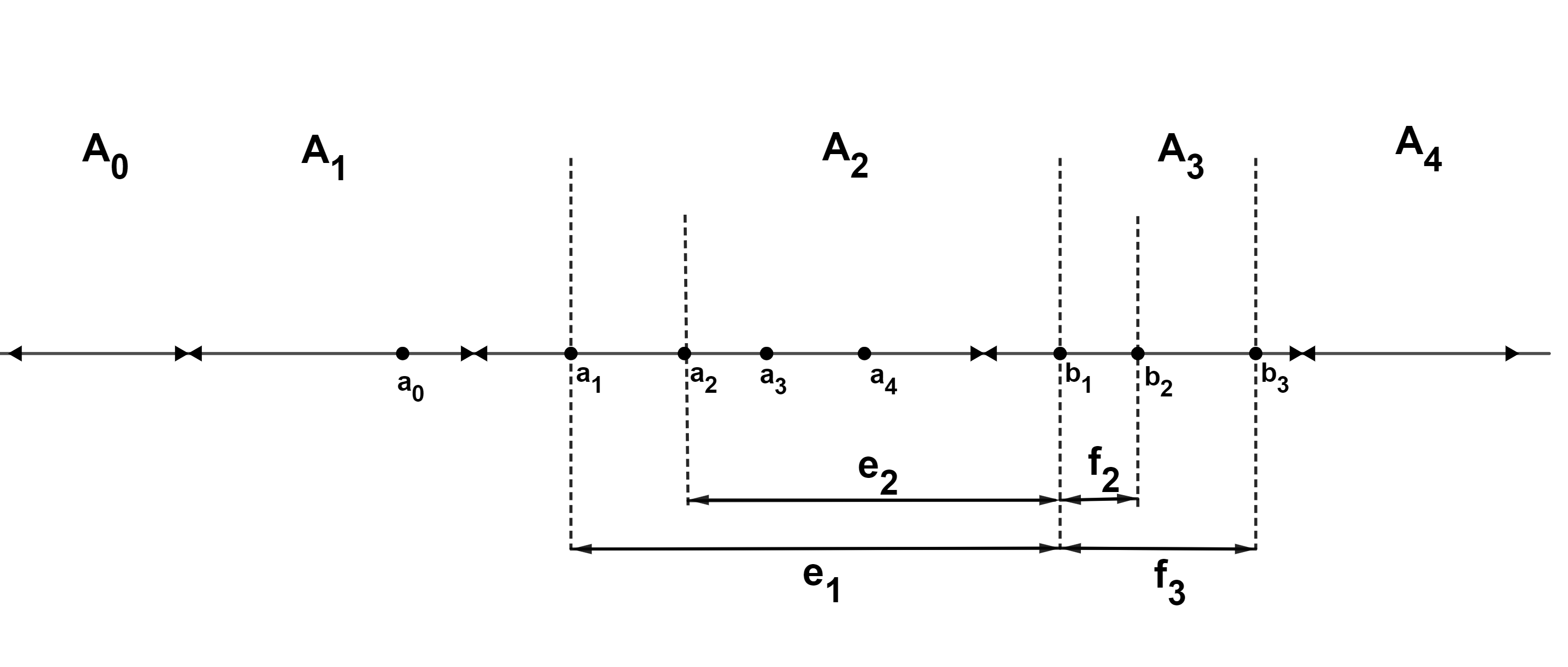}
\vspace{-0.75cm}
\caption{The notation and definitions in partitioned point set $S \cup T$.}
\label{fig:1}       
\end{figure*}

Let $A_w=\{a_1,a_2,\dots,a_s\}$ with $a_1< a_2<\dots<a_s$ and $A_{w+1}=\{b_1,b_2,\dots,b_t\}$ with $b_1< b_2<\dots<b_t$ (see Fig. \ref{fig:1}). We denote $\Vert b_1-a_i\Vert $ by $e_i$, and $\Vert b_i-b_1\Vert$ by $f_i$. Obviously $f_1=0$. We let both $a_0$ and $b_0$ denote the rightmost point of $A_{w-1}$. The cost of matching each point $a\in S$ to a point $b \in T$ is considered $\Vert a-b\Vert$. The cost of a matching is the sum of the costs of all matched pairs $(a,b)$.

Firstly, we briefly describe the algorithm given by Colannino et al. \cite{ColanDamian}, which computes a minimum-cost MM between two sets $S$ and $T$ (all points $p\in S\cup T$ have infinite capacity). The running time of their algorithm is $O(n \log n)$ and $O(n)$ for the unsorted and sorted point set $S\cup T$, respectively. We denote by $C(q)$ the cost of a minimum-cost MM for the set of the points $\{p\in S\cup T:p\leq q\}$. Their proposed algorithm computes $C(q)$ for all the points $q$ in $S\cup T$. Let $m$ be the largest point in $S \cup T$, then the cost of the minimum-cost MM between $S$ and $T$ is equal to $C(m)$.

\begin{figure*}
\vspace{-6cm}
\hspace{-11cm}
\includegraphics[width=2.7\textwidth]{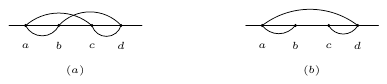}

\vspace{-36cm}
\caption{Suboptimal matchings. (a) $(a,c)$ and $(b,d)$ do not both belong to an optimal matching. (b) $(a,d)$ does not belong to an optimal matching. }
\label{fig:2}       
\end{figure*}

\newtheorem{lemma}{Lemma}
\begin{lemma}
\cite{ColanDamian} Let $b<c$ be two points in $S$, and $a<d$ be two points in $T$ such that $a\leq b<c\leq d$. Then, a minimum-cost MM that contains $(a,c)$ does not contain $(b,d)$, and vice versa (Fig. \ref{fig:2}(a)).
\label{lem3}
\end{lemma}

\begin{lemma}
\cite{ColanDamian} Let ${b,d} \in T$ and ${a,c} \in S$ with $a<b<c<d$. Then, a minimum-cost MM contains no pairs $(a,d)$ (Fig. \ref{fig:2}(b)).
\label{lem1}
\end{lemma}


\newtheorem{corollary}{Corollary}
\begin{corollary}{}
Let $M$ be a minimum-cost MM. For any matched pair $(a,d) \in M$ with $a < d$, we have $a \in A_i$ and $d \in A_{i+1}$ for some $i \geq 0$.
\label{consecutive}
\end{corollary}

\begin{lemma}{}
\cite{ColanDamian} In a minimum-cost MM, each $A_i $ for all $i>0$ contains a point $q_i$, such that all points $a \in A_i$ with $a<q_i$ are matched to the points in $A_{i-1}$ and all points $a' \in A_i$ with $q_i<a'$ are matched to the points in $A_{i+1}$.
\label{lem4}
\end{lemma}

The point $q_i$ defined in Lemma \ref{lem4} is called a \textit{separating point}. Their algorithm aims to explore the separating point of each partition $A_i$ for all $i> 0$. They assumed that initially $C(p)=\infty$, for all points $p \in A_0$. Let $A_w=\{a_1,a_2,\dots,a_s\}$ and $A_{w+1}=\{b_1,b_2,\dots,b_t\}$. Their dynamic programming algorithm computes $C(b_i)$ for each $b_i \in A_{w+1}$, assuming that $C(p)$ has been computed for all points $p < b_i$ in $S\cup T$. Depending on the values of $w$, $s$, and $i$, there are five possible cases.

\begin{figure*}
\vspace{1cm}
\hspace{1cm}
 \includegraphics[width=0.76\textwidth]{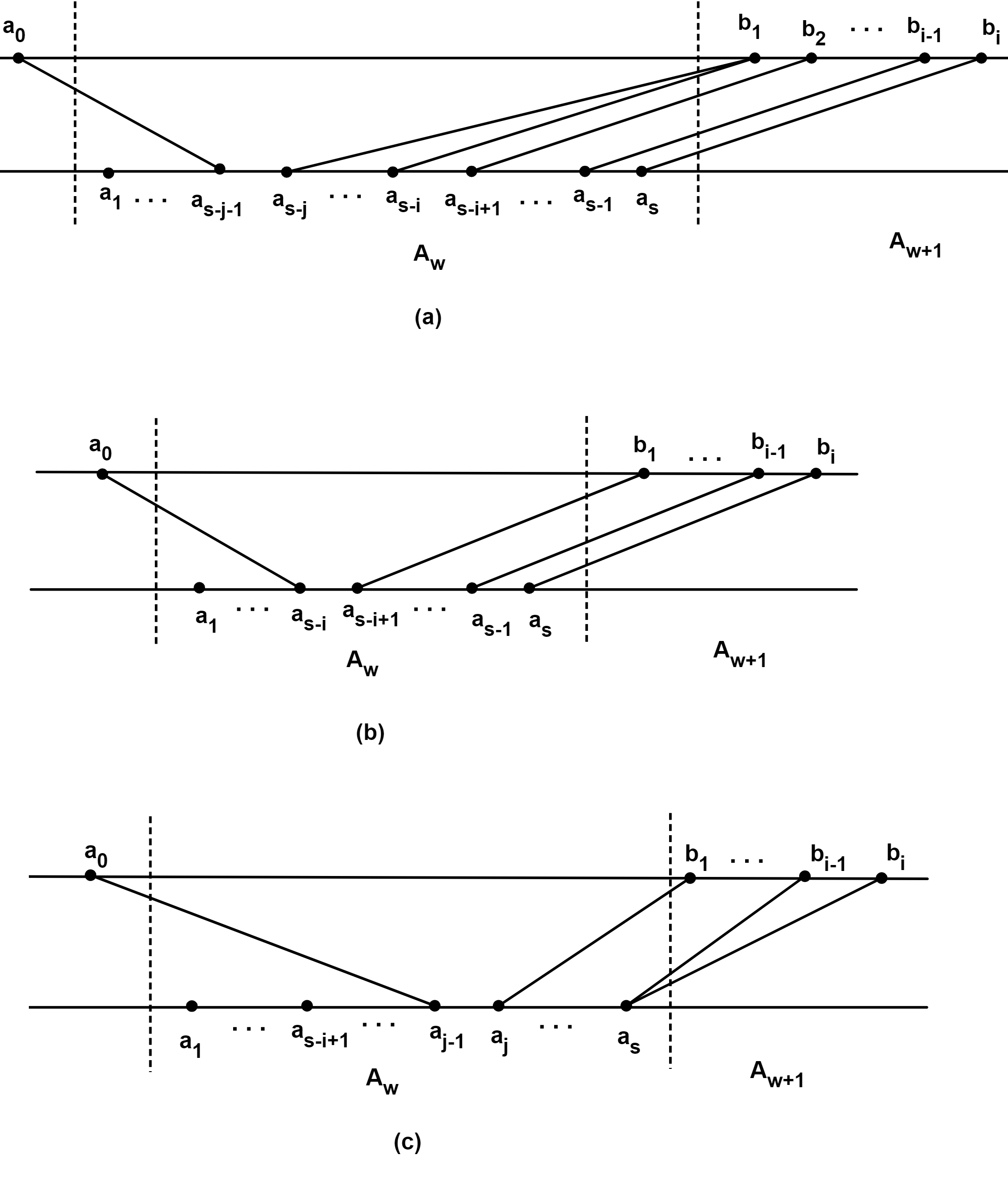}
\vspace{0cm}
\caption{Case 4: $w>0,s>1,t>1$. (a) Computing $X(b_i)$. (b) Computing $Y(b_i)$. (c) Computing $Z(b_i)$.}
\label{fig:5}       
\end{figure*}

\begin{description}
\item[Case 0:] $w=0$. In this case, there are two possible situations:
\begin{itemize}
\item $i\leq s$. We compute the optimal matching by assigning the first $s-i$ elements of $A_0$ to $b_1$ and the remaining $i$ elements pairwise (Fig. \ref{fig:3}(a)). Thus, we have:
\[C\left(b_i\right)=\sum^s_{j=1}{e_j}+\sum^i_{j=1}{f_j}.\]

\item $i>s$. The cost is minimized by matching the first $s$ points in $A_1$ pairwise with the points in $A_0$, and the remaining $i-s$ points in $A_1$ with $a_s$ (Fig. \ref{fig:3}(b)). Thus, we have:
\[C\left(b_i\right)=\left(i-s\right)e_s+\sum^s_{j=1}{e_j}+\sum^i_{j=1}{f_j}.\]

\end{itemize}


\begin{figure*}
\vspace{1.2cm}
\hspace{1cm}
\includegraphics[width=0.6\textwidth]{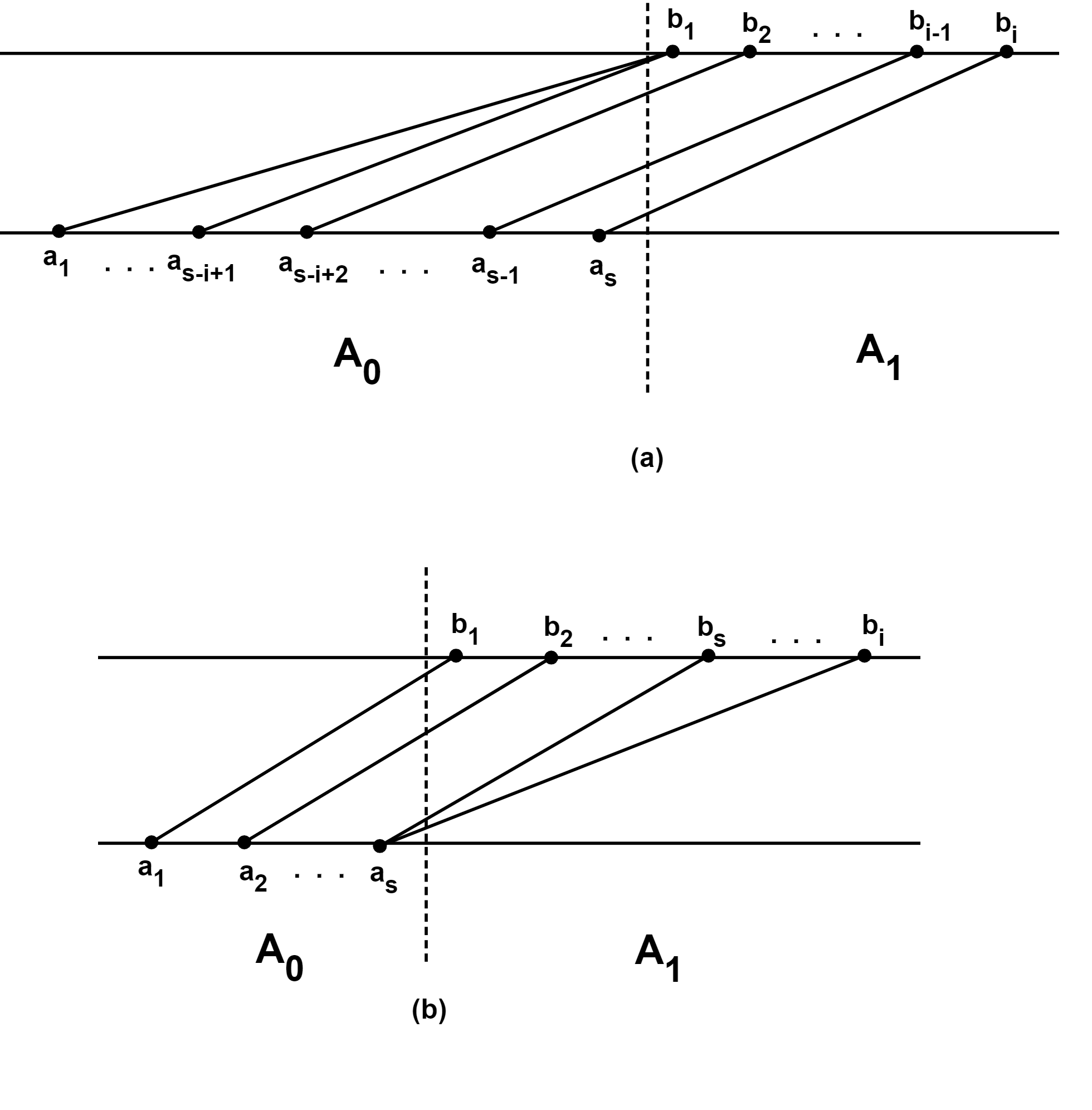}
\vspace{-0.3cm}
\caption{Case 0: $w=0$. (a) $1 \leq i \leq s$. (b) $s<i \leq t$.}
\label{fig:3}       
\end{figure*}

\item[Case 1:] $w>0,s=t=1$. Fig. \ref{fig:4}(a) illustrates this case. By Lemma \ref{lem4}, $b_1$ must be matched to the point $a_1$. Therefore, we can omit the point $a_1$, unless it reduces the cost of $C\left(b_1\right)$:
    \[C\left(b_1\right)=e_1 +\min(C(a_0),C(a_1)).\]

\item[Case 2:] $w>0,s=1,t>1$. By Lemma \ref{lem4}, we can minimize the cost of the MM by matching all points in $A_{w+1}$ to $a_1$ as presented in Fig. \ref{fig:4}(b). As Case 1, $C\left(b_i\right)$ includes $C\left(a_1\right)$ if $a_1$ covers other points in $A_{w-1}$; otherwise, $C\left(b_i\right)$ includes $C\left(a_0\right)$. Thus:
    \[C\left(b_i\right)=\sum^i_{j=1}{f_j}+ie_1+\min(C(a_0),C(a_1)).\]

\item[Case 3:] $w>0,s>1,t=1$. By Lemma \ref{lem4}, we should find the point $q_w\in A_w$ such that all points $a_j \in A_w$ with $a_j<q_w$ are matched to the points in $A_{w-1}$ and all points $a_k \in A_w$ with $q_w<a_k$ are matched to the points in $A_{w+1}$ (Fig. \ref{fig:4}(c)). Thus:
    \[C\left(b_1\right)=\min^s_{i=1}(\sum^s_{j=i}{e_j}+C(a_{i-1})).\]


\begin{figure*}
\vspace{1cm}
\hspace{3cm}
 \includegraphics[width=0.5\textwidth]{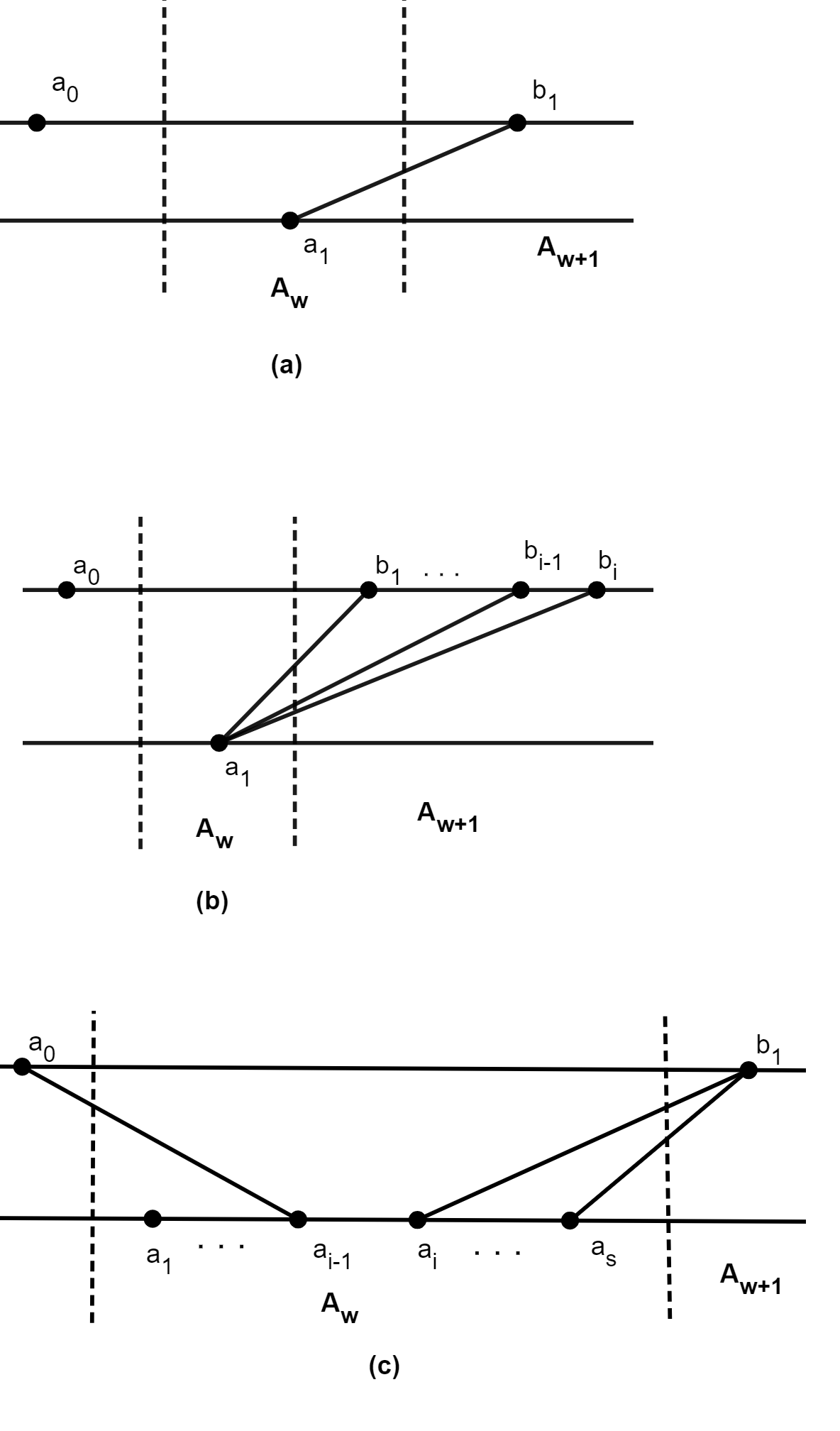}
\vspace{-0.1cm}
\caption{(a). Case 1: $w>0,s=t=1$. (b) Case 2: $w>0,s>1,t=1$. (c) Case 3: $w>0,s>1,t=1$.}
\label{fig:4}       
\end{figure*}

\item[Case 4:] $w>0,s>1,t>1$. In this case, as Case 3, we should find the point $q_w$ that splits $A_w$ to the left and right. Let $X\left(b_i\right)$ be the cost of matching $b_1,b_2,\dots,b_i$ to at least $i+1$ points in $A_w$ (Fig. \ref{fig:5}(a)). Let $Y\left(b_i\right)$ be the cost of matching $b_1,b_2,\dots,b_i$ to exactly $i$ points in $A_w$ (Fig. \ref{fig:5}(b)). Finally, let $Z\left(b_i\right)$ denote the cost of matching $b_1,b_2,\dots,b_i$ to fewer than $i$ points in $A_w$, as depicted in Fig. \ref{fig:5}(c).

    Let $S_i=\sum_{j=i}^{s}{e_j}+C(a_{i-1})$ for $i=1,2,\dots,s$ and $M_i=\min\{S_j:1\leq j \leq i\}$. Then, we have $X(b_i)= M_{s-i}+\sum_{j=1}^{i}{f_j}$ for $1\leq i<s$.
    It is easy to see that the values $X(b_i)$ and $Y(b_i)$ can be computed in $O(s + t)$ time. Also, we observe that
    $$Z(b_i) = e_s + f_i + \min(Y(b_{i-1}),Z(b_{i-1})).$$
    Therefore, we have:

\[C\left(b_i\right)\vspace{0.3cm}=\left\{
\begin{array}{lr}

\min(X\left(b_i\right),Y\left(b_i\right),Z\left(b_i\right))

 & 1\leq i \leq \min(s,t)
 \\
 \min(Y\left(b_s\right),Z\left(b_s\right)) & i=s
 \\
 C\left(b_{i-1}\right)+e_s+f_i & \min(s,t)<i\leq t
 \end{array}
\right..\]

\end{description}


\section{Our algorithms}
\label{intro}

In this section, we first present our linear time dynamic programming algorithm with a conceptually simple proof for computing a minimum-cost MM between two point sets on a line. Then, we use the idea of our first algorithm to give a linear time algorithm for finding a minimum-cost OLCMM. Note that we assume the points in $S\cup T$ are in increasing order, thus we do not consider the time for sorting $S \cup T$ in our algorithms.

\subsection{Our MM algorithm in one dimension}
\label{MM}
We use the notations and definitions used in Section \ref{PreliminSect} in our algorithm. Given a point $p \in S \cup T$, we denote by $deg(p)$ the number of points matched to $p$ in the MM. Firstly, we assume that $C(p)= \infty$ for all $p \in S \cup T$.

\begin{lemma}{}
In a minimum-cost MM denoted by $M'$, if we have $deg(p)>1$ for $p \in S\cup T$, then the cost of $M'$ includes $C(p)$, i.e. $p$ decreases it.
\label{lemma_main}
\end{lemma}

\begin{proof}
Assume for a contradiction that the point $p$ with $deg(p)>1$ does not decrease the cost of $M'$. Let $nearest(q)$ be the nearest point in $S \cup T$ to $q$ that can be matched to $q$ for any point $q\in S\cup T$. Then, if we replace each pair $(p,q)$ in $M'$ with $(nearest(q),q)$ except for one arbitrary pair, we get an MM with a smaller cost, contradicting the fact that $M'$ is a minimum-cost MM.
\end{proof}

As stated in the following observation, in a minimum-cost MM, only the first and last points of each partition might be matched to more than one point and, by Lemma \ref{lemma_main}, might decrease the cost of the MM.

\newtheorem{observation}{Observation}
\begin{observation}{}
For a point $a_k\in A_w$, if we have $deg(a_k)>1$, then:
      \begin{itemize}
        \item either $k=1<s$ ($a_1$ has been matched to more than one point of $A_{w-1}$),
        \item or $1<k=s$ ($a_s$ has been matched to more than one point of $A_{w+1}$),
        \item or $k=s=1$ ($a_1$ has been matched to more than one point of $A_{w-1}$, or more than one point of $A_{w+1}$, or at least one point of $A_{w-1}$ and one point of $A_{w+1}$).
      \end{itemize}
\label{observation1}
\end{observation}

In the following, we precisely describe our algorithm for finding an optimal MM between $S$ and $T$. Let $MM(q)$ be a minimum-cost MM for the set of the points $\{p\in S\cup T:p\leq q\}$. Recall that the capacities of the points in an MM are infinite. Assuming that we have computed $C(p)$ for all $p\leq a_s$, now we compute $C(b_1),C(b_2),\dots, C(b_t)$, respectively as follows. Firstly, we check that whether covering the points of $A_w=\{a_1,a_2,\dots,a_s\}$ by $b_1$ decreases the cost of the MM or not.

\begin{lemma}
Let $a_{j'}<a_j$ be two points in $A_{w}$. If $C(a_{j-1})+e_j \geq C(a_j)$, implying that covering $a_j$ by $b_1$ does not decrease the cost of the MM (and therefore $C(b_1)$ includes $C(a_j)$), then covering $a_{j'}$ by $b_1$ does not decrease the cost of the MM (and $C(b_1)$ includes $C(a_{j'})$), too. Thus, we have $C(a_{j'-1})+e_{j'} \geq C(a_{j'})$.
\label{lemmanew5}
\end{lemma}

\begin{proof}
We have two cases:
\begin{itemize}
  \item either $deg(a_{j'})>1$. Then, by Lemma \ref{lemma_main}, $C(b_1)$ includes $C(a_{j'})$.
  \item or $deg(a_{j'})=1$. Then, as a contradiction assume $C(a_{j'-1})+e_{j'} < C(a_{j'})$. Thus, we cover $a_{j'}$ by $b_1$; we match $a_{j'}$ to $b_1$ and remove the pair $(q',a_{j'})$, where $q'$ is the point that has been matched to $a_{j'}$. According to Lemma \ref{lem3}, two pairs $(q,a_j)$ and $(a_{j'},b_1)$ contradict the optimality, where $q$ is one of the points that have been matched to $a_{j}$. Observe that we have $q<a_{j'}<a_{j}<b_1$.
\end{itemize}
\end{proof}

Starting from $a_s$, we examine the points $a_s,a_{s-1},\dots,a_1$, respectively until $C(a_{j-1})+e_j \geq C(a_j)$ for $1\leq j \leq s$. For each $a_j \in A_w$, two cases arise:

\begin{itemize}
  \item if we have $C(a_{j-1})+e_j<C(a_j)$, which implies that covering $a_j$ by $b_1$ decreases the cost of the MM (and therefore $C(b_1)$ does not include $C(a_j)$). Then, we match $a_j$ to $b_1$ and remove the pair $(q,a_j)$ from the MM, where $q$ is the point that has been matched to $a_{j}$. Obviously, $deg(a_{j})=1$. Then, two subcases arise:

\begin{itemize}
  \item either $j=s$, implying that $deg(b_1)=0$. Then, we have:
  $$C(b_1)=C(a_{s-1})+e_s,$$

  and
$$MM(b_1)=MM(a_{s-1}) \cup (a_s,b_1).$$

  \item or $j<s$, which by Lemma \ref{lemmanew5} implies that $deg(b_1)\geq 1$. Then,
  $$C(b_1)=C(b_1)-C(a_j)+C(a_{j-1})+e_j,$$

  and

\begin{equation}
\nonumber
\begin{split}
MM(b_1)=&(MM(b_{1})\setminus MM(a_j))\\
&\cup MM(a_{j-1})\cup (a_j,b_1).
\end{split}
\end{equation}


\end{itemize}

\item if we have $C(a_{j-1})+e_j \geq C(a_j)$ for $1\leq j \leq s$, by Lemma \ref{lemmanew5}, we stop examining the points $a_{j'}<a_j$ in $A_w$.

\end{itemize}

Now, if $deg(b_1)\geq 1$, we have done; otherwise we match $b_1$ to $a_s$ as follows:
\begin{itemize}
  \item if $deg(a_s)=1$, then we check that whether $C(b_1)$ includes $C(a_s)$ or not:
$$C(b_1)=e_s+\min(C(a_s),C(a_{s-1})).$$

Thus, if $C(a_s)>C(a_{s-1})$, we have
$$MM(b_1)=MM(a_{s-1}) \cup (a_s,b_1),$$
otherwise,
$$MM(b_1)=MM(a_s) \cup (a_s,b_1).$$

 \item if $deg(a_s)>1$, by Lemma \ref{lemma_main}, $a_s$ decreases the cost of the MM, thus $C(b_1)$ includes $C(a_s)$. Then, we have:
$$C(b_1)=e_s+C(a_s),$$
and also,
$$MM(b_1)=MM(a_s) \cup (a_s,b_1).$$
\end{itemize}

Now, using the following lemma, we compute $C(b_i)$ for $i=2,3,\dots,t$, respectively.

\begin{lemma}{}
Given $A_w=\{a_1,a_2,\dots,a_s\}$ with $a_1< a_2<\dots<a_s$ and $A_{w+1}=\{b_1,b_2,\dots,b_t\}$ with $b_1< b_2<\dots<b_t$. In a minimum-cost MM, for $i \geq 2$ and $w\geq 0$, we have:


\[C(b_i)= \left\{
\begin{array}{ll}

C(b_{i-1})+f_i-f_1  &deg(b_1)>1
 \\

C(b_{i-1})+e_s+f_i &deg(a_s)>1
 \\
C'_i &
\begin{array}{l}
s-i+1>0,\\deg(a_{s-i+1})=1
\end{array}

 \\

C(b_{i-1})+e_s+f_i &  \mbox{otherwise}

\end{array}
\right.
\]
where \[
\begin{array}{ll}
C'_i= & C(b_{i-1})+f_i \\
& + \min(e_{s},e_{s-i+1}-C(a_{s-i+1})+C(a_{s-i})).
\end{array}\]

\label{lemma-new}
\end{lemma}

\begin{proof}
For each $b_i \in \{b_2, \dots,b_t\}$, we distinguish two cases:

\begin{itemize}
\item [Case 1.]$deg(b_1)>1$. Assume $b_1$ has been matched to $a_h,a_{h+1}, \dots ,a_s$.

\newtheorem{Claim}{Claim}
\begin{Claim}{}
\label{claim0}
If we remove one of the pairs $(a_k,b_1)$ for an arbitrary $k\in \{h,\dots,s\}$ from $MM(b_{i-1})$ and add $(a_k,b_i)$ to $MM(b_{i-1})$, we get $MM(b_i)$ for $i\geq 2$.
\end{Claim}

\begin{proof}
Assume for a contradiction that we do not replace the pair $(a_k,b_1)$ with the pair $(a_k,b_i)$, i.e. $b_i$ is matched to $a_s$ in the best case, while $a_h,a_{h+1}, \dots ,a_s$ have been matched to $b_1$. Then, by removing $(a_k,b_1)$ and $(a_s,b_i)$ for $k\in \{h,\dots,s\}$ and adding $(a_k,b_i)$, we get an MM with a smaller cost. Contradiction.
\end{proof}

Thus, in this case we have:
    $$C(b_i)=C(b_{i-1})+f_i-f_1.$$

Observe that removing the pair $(a_k,b_1)$ does not contradict Lemma \ref{lemma_main}, since the MM still includes $C(b_{i-1})$.

So,
$$MM(b_i)=MM(b_{i-1})\setminus (a_k,b_1) \cup (a_k,b_i).$$

\item [Case 2.]$deg(b_1)=1 $. In this case, one of the following subcases arises:

\begin{enumerate}

\item [Subcase 2.1.]$deg(a_s)>1$. Then, we observe that one of the following subcases holds:
\begin{itemize}
  \item [Subcase 2.1.1.]$i=2$. Then, we use the following claim.

\begin{Claim}{}
\label{claim1}
In this subcase, we have $s=1$.
\end{Claim}

\begin{proof}
We have $i=2$, which implies that at least one point of $A_{w-1}$ has been matched to $a_s$, since $deg(a_s)>1$ and $a_s$ has been matched to at most a single point of $A_{w+1}$ (i.e. $b_1$). Thus, by Observation \ref{observation1}, $s=1.$
\end{proof}

We have $s=1$. Moreover, from Lemma \ref{lemma_main}, $C(b_1)$ includes $C(a_1)$. Thus, in this subcase, we simply match $b_2$ to $a_s$ or actually $a_1$. Then, we have:
  $$C(b_i)=C(b_{i-1})+f_i+e_1.$$
And also,
$$MM(b_i)=MM(b_{i-1})\cup (a_1,b_i).$$

\item [Subcase 2.1.2.]$i>2$. In this subcase:
    \begin{itemize}
      \item either $s=1$. Then, according to Lemma \ref{lem4}, $b_i$ is matched to $a_s$. Thus,
      $$C(b_i)=f_i+e_1+C(b_{i-1}),$$
      and
      $$MM(b_i)=MM(b_{i-1}) \cup (a_1,b_i).$$

      \item or $s>1$. Note that by Observation \ref{observation1}, $a_s$ has been matched to no points of $A_{w-1}$ (since $s>1$). Observe that at least two points of $A_{w+1}$ (i.e. $b_1$ and $b_{i-1}$) has been matched to $a_s$, since $deg(a_s)>1$.

\begin{lemma}{}
Suppose in $MM(b_{i-1})$ we have $deg(b_1)=1$ and $deg(a_s)>1$. If $b_{i-1}$ has been matched to $a_s$, then all the points $b_{i} ,\dots, b_t$ would also be matched to $a_s$ for $i\geq 3$.
\label{cccc}
\end{lemma}

\begin{proof}
Observe that $b_{i-1}$ has been matched to $a_s$ instead of $a_k< a_s$, so we have:

\begin{equation}
\nonumber
\begin{split}
   C(b_{i-2})+e_s+f_{i-1}< & C(b_{i-2})+f_{i-1}
   \\&+e_k+C(a_{k-1})\\&-C(a_{k}),
\end{split}
\end{equation}

and thus:

$$e_s<e_k+C(a_{k-1})-C(a_{k}),$$

If we add $C(b_{i-1})$ and $f_{i}$ to both sides of the above inequality, then we have:
\begin{equation}
\nonumber
\begin{split}
C(b_{i-1})+e_s+f_{i}<&C(b_{i-1})+f_{i}\\
  &+e_k+C(a_{k-1})\\&-C(a_{k}),
\end{split}
\end{equation}

thus $a_s$ is also the optimal point for $b_{i}$.
\end{proof}

So, in this subcase it holds:
 $$C(b_i)=C(b_{i-1})+f_i+e_s,$$
 and
$$MM(b_i)=MM(b_{i-1}) \cup (a_s,b_i).$$

    \end{itemize}

\end{itemize}

\item [Subcase 2.2.]$deg(a_s)=1$. In this subcase, we observe that the $i-1$ points $b_1,b_2,\dots,b_{i-1}$ have been matched pairwise to the points $a_s,a_{s-1}, \dots, a_{s-i+2}$. We use the following claim.

\begin{Claim}{}
Let $a_k$ be the largest point in $A_w$ that has been matched to exactly one point of $A_{w-1}$ and $deg(a_k)=1$. Let $a_{k'}$ be another point in $A_w$ that has been matched to a point of $A_{w-1}$ such that $deg(a_{k'})=1$ and $a_{k'}< a_k$. Then, $b_i$ can be matched to either $a_k$ or $a_s$ but not to $a_{k'}$.
\label{claimnew}
\end{Claim}
\begin{proof}
Assume that $a_k$ has been matched to exactly one point $p\in A_{w-1}$. Suppose by contradiction that $b_i$ is matched to $a_{k'}$. Then, by Lemma \ref{lem3}, two pairs $(a_{k'},b_i)$ and $(p,a_k)$ contradict the optimality.
\end{proof}

We distinguish two subcases:

\begin{itemize}

\item [Subcase 2.2.1.]$s-i+1>0$. Then, one of the following statements holds:
\begin{itemize}

  \item $deg(a_{s-i+1})=1$. Observe that $a_{s-i+1}$ is the largest point of $A_w$ that has been matched to a point of $A_{w-1}$, thus according to Claim \ref{claimnew}, we need to check that whether $b_i$ must be matched to $a_s$ or $a_{s-i+1}$:
\begin{equation}
\nonumber
\begin{array}{l}
 C(b_{i})=\min(C(b_{i-1})+e_{s}+f_i,\\
C(b_{i-1})+e_{s-i+1}+f_i-C(a_{s-i+1})\\+C(a_{s-i})).
\end{array}
\end{equation}

Thus,
if the following holds: 

 \begin{equation}
    \nonumber
    \begin{split}
    C(b_{i-1})+e_{s}+f_i>&C(b_{i-1})+e_{s-i+1}\\
    &+f_i-C(a_{s-i+1})\\&+C(a_{s-i}),
    \end{split}
    \end{equation}

then we have:

 \begin{equation}
    \nonumber
    \begin{split}
    MM(b_i)=&(MM(b_{i-1})\setminus MM(a_{s-i+1}))\\
    & \cup MM(a_{s-i})\cup (a_{s-i+1},b_i),
        \end{split}
    \end{equation}

otherwise,
$$MM(b_i)=MM(b_{i-1}) \cup (a_s,b_i).$$

\item $deg(a_{s-i+1})>1$, which by Observation \ref{observation1} implies that $s-i+1=1$. Thus, by Lemma \ref{lemma_main} and Claim \ref{claimnew}, $b_i$ is matched to $a_s$:
    $$C(b_i)=C(b_{i-1})+f_i+e_s.$$
Thus,
$$MM(b_i)=MM(b_{i-1}) \cup (a_s,b_i).$$

\end{itemize}

\item [Subcase 2.2.2]$s-i+1=0$. This subcase implies that all the points in $A_w$ have been matched to the points $b_1,b_2,\dots,b_{i-1}$ pairwise. So, there does not exist any point $a_k$ in $A_{w}$ with $deg(a_k)=1$ that has been matched to only a single smaller point. Thus, by Claim \ref{claimnew}, we simply match $b_i$ to the closest point in $A_w$, i.e., $a_s$. Therefore,
$$C(b_i)=C(b_{i-1})+f_i+e_s,$$
and
$$MM(b_i)=MM(b_{i-1}) \cup (a_s,b_i).$$


\end{itemize}

\end{enumerate}

\end{itemize}\end{proof}

\begin{theorem}{}
Given two point sets $S$ and $T$ on a line with $n=\vert S\vert+\vert T\vert$, our algorithm computes a minimum-cost MM between $S$ and $T$ in $O(n)$ time.
\end{theorem}

\subsection{Our OLCMM algorithm}
\label{OLCCMMal}
In this section, we give the first linear time algorithm tackling the OLCMM problem faster than the $O(n^2)$ algorithm presented in \cite{Rajabi-Alni} (Algorithm \ref{OLCMM}). Initially, we let $M=\emptyset$ (Line 1 of Algorithm \ref{OLCMM}). Observe that if $\sum_{v\in S}Cap(v)<\vert T\vert$ or $\sum_{v\in T}Cap(v)<\vert S\vert$, then there does not exist an OLCMM between $S$ and $T$ with the capacity constraint $Cap(v)$ for each $v \in S\cup T$ (Lines 2--4 of Algorithm \ref{OLCMM}). Our algorithm is based on the dynamic programming algorithm described in Section \ref{MM}: $C(p,j)$ denotes the cost of a minimum-cost OLCMM between the points $\{q\in S \cup T: q\leq p\}$ such that the capacity of each point $q<p$ is equal to $Cap(q)$, i.e. its actual capacity, but $Cap(p)=j$.

Let $MM(p,j)$ denote the OLCMM corresponding to $C(p,j)$. We examine the points $b_1,b_2, \dots, b_t$ of each partition $A_{w+1}$ for $w\geq0$, respectively, and compute $C(b_i,j)$ (by finding $MM(b_i,j)$), for $j=1,2, \dots, Cap(b_i)$ if needed (Lines 11--29 of Algorithm \ref{OLCMM}). Let $MA(p,k)$ be the point that occupies the $k$th capacity of $p$. Note that we can use linked lists for maintaining $MM(p,k)$ and $C(p,k)$. We assume that if $MM(p,k)$ and $C(p,k)$ do not exist in the linked lists (i.e. $MM(p,k)=null$ and $C(p,k)=null$), then $MM(p,k)=\emptyset$ and $C(p,k)= \infty$ for all $p \in S \cup T$ and $1\leq k \leq Cap(p)$. Obviously, the number of pairs in a minimum-cost OLCMM is equal to $\max(\vert T\vert ,\vert S\vert)$. We assume that $C(b_i,0)=C(b_{i-1},deg(b_{i-1}))$ and $C(b_i,0)=C(a_s,deg(a_s))$ for $i>1$ and $i=1$, respectively.

\begin{theorem}{}
Let $S$ and $T$ be two point sets on a line with total cardinality $n$, then we can compute a minimum-cost OLCMM between $S$ and $T$ in linear time.
\label{theorem1}
\end{theorem}

\begin{proof}
We start with a useful corollary and lemma. We observe that Lemma \ref{lem3} also holds in a minimum-cost OLCMM.


\begin{corollary}{}
\cite{Rajabi-Alni} In a minimum-cost OLCMM, each partition $A_w$ has a point $q_w$, called the \textit{separating point} of $A_w$, such that all points $a\in A_w$ with $a\leq q_w$ are matched to some points $b\in S\cup T$ with $b\leq q_w$ and all points $a'\in A_w$ with $a'\geq q_w$ are matched to some points $b'\in S\cup T$ with $b'\geq q_w$ for $w\geq0$.
\label{consecutive2}
\end{corollary}

\begin{lemma}{}
In a minimum-cost OLCMM, a point $p \in S\cup T$ decreases the cost of the OLCMM, i.e. the cost of the OLCMM includes $C(p,1),C(p,2),\dots,C(p,deg(p))$, if $deg(p)>1$.
\label{lem3-3}
\end{lemma}

\begin{proof}
The proof of this lemma is the same as Lemma \ref{lemma_main}.
\end{proof}

Now, our algorithm can be formulated as follows. Starting from $A_1$, we examine the points $b_1, b_2, \dots,b_t \in A_{w+1}$ for each $w\geq 0$, and compute $C(b_i,j)$ for $j=1,2,\dots,Cap(b_i)$ if it is necessary.

Suppose that we have examined all points $p\in S \cup T$ with $p < b_i$ and computed $C(p,k)$ for $k=1,2,\dots,j$, respectively where $1\leq j\leq Cap(p)$. Note that in practice it is not necessary to compute $C(p,j)$ for all $j=1,2,\dots,Cap(p)$. In the following, we describe that how we compute $C(b_i,k)$ for $k=1,2,\dots,j$, respectively where $1\leq j\leq Cap(b_i)$. We execute two steps for the point $b_i$ (Lines 12, 19, and 23 of Algorithm \ref{OLCMM}).


\algsetblock[Name]{Initial}{}{3}{1cm}
\alglanguage{pseudocode}
\begin{algorithm2e}[h]

\caption{OLCMM AlgorithMa($S$,$T$)}
\label{OLCMM}

\SetAlgoVlined

\SetKwInOut{Input}{input}
\SetKwInOut{Output}{output}
\Input{Two sets $S$ and $T$ on the line with total cardinality $n$ and the capacity $Cap(v)$ for each $v\in S\cup T$}
\Output{An OLCMM between $S$ and $T$}

 $M=\emptyset$\;
\If{$(\sum_{v\in S}Cap(v)<\vert T\vert) \lor (\sum_{v\in T}Cap(v)<\vert S \vert)$}{
  Print "There does not exist an OLCMM between $S$ and $T$ with the capacity constraint $Cap(v)$ for each $v \in S\cup T$"\;

   \Return $M$\;}

\Else{
    Partition $S \cup T$ into maximal subsets $A_0,A_1, \dots $ alternating between subsets in $S$ and $T$\;
    $w=0$\;

   \While{$\vert M\vert<\max(\vert T\vert,\vert S\vert)$}{
       $A_w=\{a_1,a_2, \dots, a_s\}$, $A_{w+1}=\{b_1,b_2, \dots, b_t\}$, and $a_0=b_0=\max(p \in A_{w-1})$\;
       $ns(w+1)=w+1$, $next1(w)=a_s$, $next2(w)=a_s$, and $id(w+1)=0$, $continue(w+1)=True$\;
      \For {$i=1$ to $t$}{
         \Call{Step1}{$M$, $MM$, $S$, $T$, $i$, $A_0,A_1, \dots$}\;
      \lIf{$deg(b_i)>1$}{
          $id(w+1)=i$}
    \ElseIf{$deg(b_i)=0$}{
          $w'=ns(w+1)$\;
        \While{$(deg(b_i)<1)\land (w'\geq 0)$}{
           \If {$((w+1-w')\mod 2)=0$}{
             \If{$w'\geq 1$}{
                  \Call{Step2.1}{$M$, $MM$, $S$, $T$, $i$, $A_0,A_1, \dots$,$w'$}\;}
              \lIf{$deg(b_i)<1$}{
                   $w'=w'-1$}
             }

           \If{$w'\geq 0$}{
           \If{$deg(b_i)<1$}{
               \Call{Step2.2}{$M$, $MM$, $S$, $T$, $i$, $A_0,A_1, \dots$,$w'$}\;}
           \If{$deg(b_i)<1$}{
               $w'=w'-1$\;
           \If{$w'\geq 1$}{
               $w'=ns(w')$\;

              }

              \lElse{
               $w'=-1$
              }

              }
}

        }%
}%

    $ns(w+1)=w'$\;
    }%
   $w=w+1$\;
}%
  \Return $M$\;
}
 \end{algorithm2e}

\begin{itemize}
\item [Step 1.]In this step, we should check that whether covering the points $q<b_i$ by $b_i$ decreases the cost of the OLCMM or not as described in Algorithm \ref{Case1.1} (i.e., by Corollary \ref{consecutive2}, we should find the separating points of some previous partitions).

\begin{Claim}{}
Given two points $a_j,a_k$ with $a_k< a_j<b_i$, if one of the following properties holds:
\begin{itemize}
  \item $deg(a_j)>1$,
  \item or $deg(a_j)=1$, $MA(a_j,1)\leq a_j$, and $$C(a_j,1)\leq C(a_{j-1},deg(a_{j-1}))+b_i-a_j,$$

\end{itemize}

then,
$$C(a_k,deg(a_k))\leq C(a_k,deg(a_k)-1))+b_i-a_k,$$
which implies covering $a_k$ by $b_i$ does not decrease the cost of the OLCMM.
  \label{claim2}
  \end{Claim}

\begin{proof}
We distinguish between two cases:
  \begin{itemize}
    \item $deg(a_k)>1$. Then, by Lemma \ref{lem3-3}, the OLCMM includes $C(a_k,deg(a_k))$.


    \item $deg(a_k)=1$ and $MA(a_k,1)\leq a_k$. In this case,
    \begin{itemize}
      \item either $\{MA(a_j,h):MA(a_j,h)\leq a_j\}_{h=1}^{deg(a_j)}\neq\emptyset$. Then (by assumption),
      \begin{itemize}
        \item either $deg(a_j)>1$, then by Lemma \ref{lem3-3}, the OLCMM includes $C(a_j,h)$ for $1\leq h \leq deg(a_j)$.
        \item or  $deg(a_j)=1$, $MA(a_j,1)\leq a_j$, and $$C(a_j,1)\leq C(a_{j-1},deg(a_{j-1}))+b_i-a_j.$$
      \end{itemize}
      Let $p'\in \{MA(a_j,h):MA(a_j,h)\leq a_j\}_{h=1}^{deg(a_j)}$. In both two above cases, the OLCMM includes the pair $(a_j,p')$. Now, suppose by contradiction that covering $a_k$ by $b_i$ decreases the cost of the OLCMM, i.e.,
       $$C(a_k,1) > C(a_{k-1},deg(a_{k-1}))+b_i-a_k,$$
       and add the pair $(a_k,b_i)$ to the OLCMM and remove the pair $(MA(a_k,1),a_k)$. Then, depending on $p'$ by either Lemma \ref{lem3} or Lemma \ref{lem1}, two pairs $(a_k,b_i)$ and $(p',a_j)$ contradict the optimality.

      \item or $\{MA(a_j,h):MA(a_j,h)\leq a_j\}_{h=1}^{deg(a_j)}=\emptyset$. This case implies that by assumption $deg(a_j)>1$. Let $p$ be the largest point that has been matched to $a_j$, i.e., $p=\max(\{MA(a_j,h)\}_{h=1}^{deg(a_j)})$. Assume by contradiction that covering $a_k$ by $b_i$ decreases the cost of the matching, i.e.,
        $$C(a_k,1) > C(a_{k-1},deg(a_{k-1}))+b_i-a_k.$$

         Observe that $a_j<p\leq b_i$, thus we have:
         $$C(a_k,1)> C(a_{k-1},deg(a_{k-1}))+p-a_k.$$ Note that by assumption $a_k$ has not been matched to $p$, i.e., covering $a_k$ by $p$ does not decrease the cost of the OLCMM. Contradiction.
    \end{itemize}

  \end{itemize}
  Thus, we conclude the lemma.
\end{proof}

  In the following, we describe this step precisely. See Fig. \ref{fig:2bakeri}. Firstly, if $i=1$, we call $\Call{Proc}{b_1,A_{w},j,w+1}$ (Lines 1--3 of Algorithm \ref{Case1.1}), which starting from $a_s \in A_w$, examines the points $a_j \in \{a_s,a_{s-1}, \dots, a_1\}$, respectively until $j<1$ or $deg(b_i)=Cap(b_i)$ (Lines 1--15 of Algorithm \ref{Proc1}). For the point $a_j \in \{a_s,a_{s-1}, \dots, a_1\}$, one of the following cases arises:

  \begin{itemize}

    \item $deg(a_j)\leq 1$. In this case, we distinguish three subcases (Lines 2--14 of Algorithm \ref{Proc1}).

    \begin{itemize}

    \item $deg(a_j)=0$, which implies that $b_1$ is the first point having enough capacities for $a_j$. Thus, $a_j$ is matched to $b_1$ (Lines 3--6 of Algorithm \ref{Proc1}):
        $$C(b_1,deg(b_1))=C(b_1,deg(b_1)-1)+b_1-a_j.$$

    \item $deg(a_j)=1$ and $MA(a_j,1)\leq a_j$. Then,

    \begin{itemize}

    \item either $C(a_j,1)>C(a_{j-1},deg(a_{j-1}))+b_1-a_j$, which means that covering $a_j$ by $b_1$ decreases the cost of the OLCMM (the minimum-cost OLCMM does not include $C(a_j,1)$). Thus, we remove the pair $(a_j,MA(a_j,1))$ and add the pair $(b_1,a_j)$ (Lines 8--11 of Algorithm \ref{Proc1}):
    \begin{equation}
    \nonumber
    \begin{split}
      C(b_1,deg(b_1))=&C(b_1,deg(b_1)-1) \\&+b_1-a_j
    -C(a_j,1)\\&+C(a_{j-1},deg(a_{j-1})).
    \end{split}
    \end{equation}

    Then, we let $j=j-1$ to examine the previous point, i.e. $a_{j-1}$ (Line 12 of Algorithm \ref{Proc1}).

    \item or $C(a_j,1)\leq C(a_{j-1},deg(a_{j-1}))+b_1-a_j$, implying that covering $a_j$ by $b_1$ does not decrease the cost of the OLCMM. Then, by Claim \ref{claim2}, we let $j=0$ (and also $continute(w+1)=False$) to stop checking the points $a_{j-1}, \dots ,a_{1}$ (Line 13 of Algorithm \ref{Proc1}). Note that initially we have $continue(w+1)=True$ for all $w\geq 0$ (Line 10 of Algorithm \ref{OLCMM}).

\end{itemize}

\item $deg(a_{j})=1$ and $MA(a_{j},1)> a_{j}$. Then, we let $j=j-1$ (Line 14 of Algorithm \ref{Proc1}).

\end{itemize}

\item $deg(a_j)>1$. Then, by Claim \ref{claim2}, we let $j=0$ and $continute(w+1)=False$ to stop checking the points (Line 15 of Algorithm \ref{Proc1}).

\end{itemize}

\begin{figure*}
\vspace{1cm}
\hspace{0cm}
 \includegraphics[width=1\textwidth]{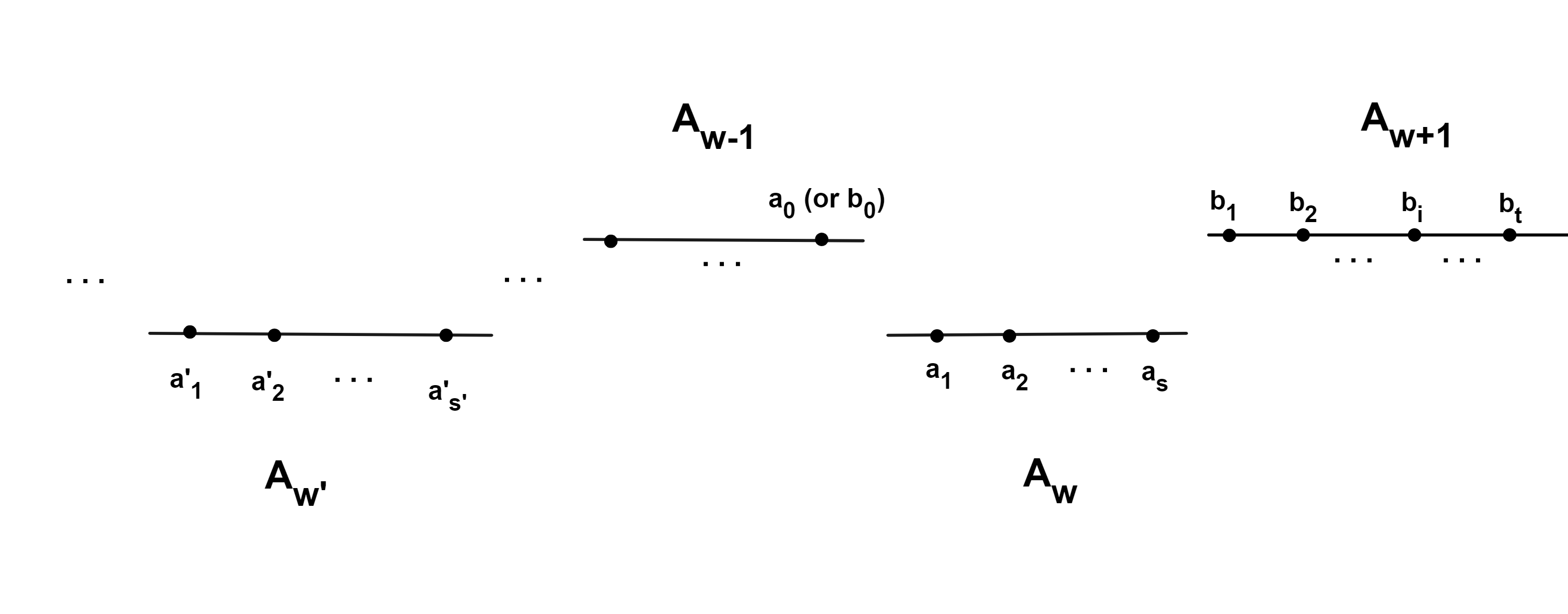}
\vspace{-1cm}
\caption{An example for illustration of Step 1.}
\label{fig:2bakeri}       
\end{figure*}

Now, let $w''=w+1$ and $t'=i-1$ (Line 4 of Algorithm \ref{Case1.1}). Suppose $a'_{s'}\in A_{w'}$ is the smallest point that has been matched to $b'_{t'}$ (Lines 6--7 of Algorithm \ref{Case1.1}). Starting from $a'_{s'-1}$, we check the points $a'_{j'}\in \{a'_{s'-1},a'_{s'-2},\dots,a'_1\}$, respectively until $deg(b_i)=Cap(b_i)$ or $j'< 1$ to verify that whether covering the points $a'_{j'}$ by $b_i$ decreases the cost of the OLCMM or not within a call to $\Call{Proc}{b_i,A_{w'},j',w''}$ (Line 9 of Algorithm \ref{Case1.1}).


For the point $a'_{j'}$, one of the following cases arises (Lines 1--15 of Algorithm \ref{Proc1}):

  \begin{itemize}

  \item $deg(a'_{j'})\leq 1$. Then, we have the following subcases (Lines 2--14 of Algorithm \ref{Proc1}):

  \begin{itemize}
    \item $deg(a'_{j'})=0$. Obviously, in this subcase, $a'_{j'}$ is matched to $b_i$ (Lines 3--6 of Algorithm \ref{Proc1}).
    \item $deg(a'_{j'})=1$ and $MA(a'_{j'},1)\leq a'_{j'}$. Then:
    \begin{itemize}
      \item if $C(a'_{j'},1)> C(a'_{j'-1},deg(a'_{j'-1}))+b_1-a'_{j'}$ (implying that covering $a'_{j'}$ by $b_i$ decreases the cost of the OLCMM), we match $b_i$ to $a'_{j'}$ and remove the pair $(a'_{j'},MA(a'_{j'},1))$ from the OLCMM (Lines 8--11 of Algorithm \ref{Proc1}). Then, we let $j'=j'-1$ (Line 12 of Algorithm \ref{Proc1}).
      \item otherwise, by Claim \ref{claim2}, we do not continue the search; we let $j'=0$ and $continute(w'')=False$ to exist from the while loops of Algorithms \ref{Proc1} and \ref{Case1.1}, respectively (Line 13 of Algorithm \ref{Proc1}).
    \end{itemize}

 \item $deg(a'_{j'})=1$ and $MA(a'_{j'},1)> a'_{j'}$. In this subcase, we let $j'=j'-1$ (Line 14 of Algorithm \ref{Proc1}).

  \end{itemize}

 \item $deg(a'_{j'})>1$. In this case, as above, we let $j'=0$ and $continute(w'')=False$. Thus, checking the points stops (Line 15 of Algorithm \ref{Proc1}).

 \end{itemize}

Now, if $continute(w'')=True$, we let $w''=w'-1$ and $t'=\vert A_{w''}\vert$ (Line 10 of Algorithm \ref{Case1.1}). While $w''\geq 1$ and $deg(b_i)<Cap(b_i)$ and $continute(w'')=True$, we do as above, iteratively (Line 5--10 of Algorithm \ref{Case1.1}). Then, if $w''< 1$ or $continue(w'')=False$, we let $continute(w+1)=False$ (Lines 11--12 of Algorithm \ref{Case1.1}).

   \begin{observation}
  If $b_i$ is matched to $a'_{j'}$ with the properties
\begin{itemize}
  \item either $deg(a'_{j'})=0$,
  \item or $deg(a'_{j'})=1$, $MA(a'_{j'},1)\leq a'_{j'}$, and $$C(a'_{j'},1)> C(a'_{j'-1},deg(a'_{j'-1}))+b_1-a'_{j'},$$
\end{itemize}

  then the pair $(a'_{j'},b_i)$ does not contradict the optimality of $C(b_{i-1},deg(b_{i-1}))$ (and $MM(b_{i-1},deg(b_{i-1}))$).
  \label{observationstep1}
  \end{observation}

  \begin{proof}
  Assume that $b_i\in T$ (resp. $b_i\in S$). It is easy to show that in $MM(b_{i-1},deg(b_{i-1}))$ all points $p\in T$ (resp. $p\in S$) with $a'_{j'}< p< b_i$ have been filled to their capacities by some points $q \in S$ (resp. $q\in T$) with $a'_{j'}<q$. Thus, there might exist only the pairs $(q,p)$ between the points $a'_{j'}$ and $b_i$ such that $a'_{j'}<q<p<b_i$. Observe that $p -q+b_i-a'_{j'}=p-a'_{j'}+b_i-q$, thus the pairs $(a'_{j'},b_i)$ and $(q,p)$ do not contradict the optimality.
  \end{proof}

  Note that, in this step, we inspect the points $q<b_i$, starting from the last point that has been examined previously. Indeed, each point $q\in S\cup T$ might not be examined twice all over our algorithm by this step. So, the overall time complexity of this step is $O(n)$.


Now, if $deg(b_i)\geq 1$ we have done, otherwise we go to Step 2.

\item [Step 2.]In this step, we seek the points $p\leq b_i$ to find an optimal point for $b_i$ to be matched such that the cost of the OLCMM is minimized. Let $ns(w+1)$ denote the last partition $A_{w'}$ that has been searched for finding enough capacities for the points of $A_{w+1}$ for $w\geq 0$. Starting from $w'=ns(w+1)$, we do the following two steps iteratively in a while loop until it holds $deg(b_i)=1$ or $w'<0$ (Lines 15--28 of Algorithm \ref{OLCMM}). Initially, we have $ns(w+1)=w+1$ for all $w\geq 0$ (Line 10 of Algorithm \ref{OLCMM}). Note that $((w+1-w')\mod 2)=0$ implies that both $A_{w'}$ and $A_{w+1}$ are subsets of $S$ ($T$) (Line 17 of Algorithm \ref{OLCMM}). Then, if $w'\geq 1$, we do Step 2.1 which is described in the following (Lines 18--19 of Algorithm \ref{OLCMM}).

 \begin{itemize}

 \item [Step 2.1.]This substep is as follows. Let $\{b'_1,b'_2,\dots,b'_{t'}\}=\{p\in A_{w'}:p< b_i\}$, which implies that $t'$ is the number of the points in $A_{w'}$ that are smaller than $b_i$ (Line 1 of Algorithm \ref{Case2}). Then:
\begin{enumerate}
  \item If $id(w')>0$, implying that there exists a point $b'_j$ in $A_{w'}$ with $1 \leq j\leq t'$ satisfying $deg(b'_{j})>1$, we do Lines 2--7 of Algorithm \ref{Case2} as follows. Recall that after we do Step 1 for each $b'_j\in A_{w'}$, if $deg(b'_j)>1$ holds, then we let $id(w')=j$ (Line 13 of Algorithm \ref{OLCMM}). Thus, $b'_{id(w')}$ is the largest point of $A_{w'}$ that has been matched to more than one point, i.e. $$MA(b'_{id(w')},1),MA(b'_{id(w')},2),\dots,MA(b'_{id(w')},deg(b'_{id(w')})),$$ denoted by $m_1,m_2,\dots,m_v$ (Line 3 of Algorithm \ref{Case2}). Observe that we have $m_h\leq b'_{id(w')}$ for all $1\leq h\leq v$ (see Observation \ref{observationnew2}). Note that initially we let $id(w')=0$ for all $w'\geq 0$ (Line 10 of Algorithm \ref{OLCMM}).

     \begin{Claim}
     \label{claimstep2}
      The minimum-cost OLCMM $MM(b_i,1)$ can be computed by matching $b_i$ to one of the points $m_1,m_2, \dots, m_v$ (arbitrarily), say $m_u$, and removing the pair $(m_u,b'_{id(w')})$.
     \end{Claim}
     \begin{proof}
     By contradiction, assume that we add a pair $(p,b_i)$ to $MM(b_i,1)$ instead of replacing the pair $(m_u,b'_{id(w')})$ with the pair $(m_u,b_i)$. Thus, $b_i$ is matched to $p$ that for which one of the following statements holds:
    \begin{itemize}
      \item $p\in \{m_1,m_2,\dots,m_v\}$. Then, we can remove the pair $(p,b'_{id(w')})$ from $MM(b_i,1)$ and get a smaller cost. Contradiction.
      \item $p\notin \{m_1,m_2,\dots,m_v\}$. We distinguish two cases: $deg(p)=1$ and $deg(p)>1$. (i) In the case where $deg(p)=1$, the statement $deg(b'_{id(w')})>1$ implies that matching the point $p$ to $b'_{id(w')}$ does not decrease the cost of $MM(b_i,1)$ (since $p$ has not been matched to $b'_{id(w')}$). Thus,

          $$C(p,1)\leq C(p,0)+b'_{id(w')}-p,$$ and therefor:
          \begin{equation}
         \nonumber
         \begin{split}
           b_i-b'_{id(w')}<&C(p,0)+b_i-b'_{id(w')} \\
          &+b'_{id(w')}-p.
         \end{split}
         \end{equation}

    \end{itemize}

    This means that the cost of matching $b_i$ to the point $p$ is larger than the cost of replacing the pair $(m_u,b'_{id(w')})$ with the pair $(m_u,b_i)$, which yields a contradiction.

   (ii) In the case where $deg(p)>1$, we have:
   $$b_i-p<b_i-b'_{id(w')},$$
  this contradicts $p< b'_{id(w')}$.
  \end{proof}
    Thus, if $id(w')>0$, we do as follows. We assume w.l.o.g that $m_u=m_v=MA(b'_{id(w')},deg(b'_{id(w')}))$, then:
     \begin{enumerate}
     \item We match $b_i$ to $m_v$ and remove $(m_v,b'_{id(w')})$ from the OLCMM (Lines 4-5 of Algorithm \ref{Case2}). Then, we let
    $$C(b_i,1)=C(b_{i-1},deg(b_{i-1}))+b_i-b'_{id(w')}.$$
    \item Next, if $deg(b'_{id(w')})=1$, we let $id(w')=id(w')-1$ (Lines 6--7 of Algorithm \ref{Case2}).
        \end{enumerate}

      \item Otherwise, if there does not exist such a point $b'_j<b_i$ with $deg(b'_j)>1$, we check the point $MA(b'_{t'},1)$ denoted by $m'$. If we have $1<deg(m')<Cap(m')$, by Lemma \ref{cc888}, we simply match $b_i$ to $m'$ (Lines 8--12 of Algorithm \ref{Case2}). Then, we have:
    $$C(b_i,1)=C(b_{i-1},1)+b_i-m'.$$

\begin{lemma}{}
In the minimum-cost OLCMM $MM(b_i,1)$, if $b'_{t'}\in A_{w'}$ has been matched to $m'$ with $1<deg(m')<Cap(m')$, then $m'$ is also the optimal point for $b_{i}\in A_{w+1}$ for $i\geq 2$, i.e. $b_{i}$ would also be matched to $m'$.
\label{cc888}
\end{lemma}

\begin{proof}
Note that, by Lemma \ref{lem3-3}, the statement $deg(m')> 1$ implies that $m'$ decreases the cost of $MM(b_i,1)$. In the minimum-cost OLCMM $MM(b_i,1)$, $b'_{t'}$ has been matched to $m'$ instead of any other point $m''$ with $deg(m'')\geq 1$, thus:

\begin{equation}
\nonumber
\begin{split}
C(b'_{t'-1},1)+b'_{t'}-m'<&C(b'_{t'-1},1)+ \min(b'_{t'}\\
 &-m''-C(m'',deg(m''))\\&+C(m'',deg(m'')\\&-1),b'_{t'}-m''),
\end{split}
\end{equation}


and thus:

\[
\begin{split}
-m'<& -m''+ \min(-C(m'',deg(m'')) \\
     &+C(m'',deg(m'')-1),0),
\end{split}\]

If we add $C(b_{i-1},1)$ and $b_i$ to both sides of the above inequality, then we have:

\begin{equation}
  \nonumber
  \begin{split}
 C(b_{i-1},1)+b_i-m'<&C(b_{i-1},1)+b_i-m''  \\
 &+ \min(-C(m'',deg(m''))\\&+C(m'',deg(m'')-1),0),
 \end{split}
 \end{equation}

thus, $b_i$ is also matched to $m'$.
\end{proof}

\item Otherwise, if $i>1$, we check as above that whether $1<deg(MA(b_{i-1},1))<Cap(MA(b_{i-1},1))$ and $b_i$ is matched to $m'=MA(b_{i-1},1)$ or not (Lines 13--18 of Algorithm \ref{Case2}).
\end{enumerate}

Then, if $deg(b_i)<1$, we let $w'=w'-1$ (Line 20 of Algorithm \ref{OLCMM}). Now, if $w'\geq 0$, we do as follows (Lines 21--28 of Algorithm \ref{OLCMM}). If we still have $deg(b_i)<1$, we do Step 2.2 which is described in the following (Lines 22-23 of Algorithm \ref{OLCMM})

\item [Step 2.2.] In this step, we consider two cases (Algorithm \ref{Case3}):

\begin{enumerate}

\item [Case 1:]If $i=1$, this means that we have computed $C(a_s,deg(a_s))$ (and $MM(a_s,deg(a_s))$), and now we want to compute $C(b_1,1)$ (and $MM(b_1,1)$ ). In this point, we observe that $a_s$ has been matched to one or more smaller points $p<a_s$, since in our dynamic programming algorithm, we insert the points of $S\cup T$ from the left to the right, respectively. 

      \begin{observation}{}
    Assume that the point $q$ with $q<p$ has been matched to at least one smaller point $p'$ (i.e. $MA(q,h)\leq q$ for some $1\leq h\leq deg(q)$ and $p'\in \{MA(q,h):MA(q,h)\leq q\}_{h=1}^{deg(q)}$). Then, in a minimum-cost OLCMM, the point $p$ can not be matched to any point $q'$ with $q'<q$ where $q,q'\in S$ (resp. $q,q'\in T$) and $p,p' \in T$ (resp. $p,p'\in T$).
 \label{observationnew2}
  \end{observation}
\begin{proof}
Suppose by contradiction that $p$ is matched to the point $q'$ with $q'<q$. Then, one of the following statements holds:
\begin{itemize}
  \item $p'<q'<q<p$. Then, by Lemma \ref{lem3}, two pairs $(q',p)$ and $(p',q)$ contradict the optimality.
  \item $q'<p'<q<p$. Then, we can simply replace two pairs $(q',p)$ and $(p',q)$ with the pairs $(q',p')$ and $(q,p)$, and get a smaller cost. Contradiction.
\end{itemize}
\end{proof}


    Thus, by Observation \ref{observationnew2}, $b_1$ can not be matched to any point $a_k$ with $a_k <a_s$. Therefor, in an optimal OLCMM, we can only match $b_1$ to $a_s$ (Lines 1--16 of Algorithm \ref{Case3}). There are two subcases:
\begin{enumerate}
  \item [Subcase 1.1:]$deg(a_s)=Cap(a_s)$ (Lines 2--4 of Algorithm \ref{Case3}). Let $b'_u\leq a_s$ be the smallest point that has been matched to $a_s$, i.e.,
  $$b'_u=\min(MA(a_s,1), \dots,MA(a_s,deg(a_s))).$$

  In this subcase, we should remove the pair $(a_s,b'_u)$, and add the pair $(a_s,b_1)$. Assume by contradiction that $b'_u$ is not the smallest point that has been matched to $a_s$, i.e., there exists a point $b'_v$ with $b'_v< b'_u$ such that $b'_v=MA(a_s,h)$ for $1\leq h\leq deg(a_s)$. Then, if we remove the pair $(b_u',a_s)$, the point $b'_u$ might be matched to a point $a'_{s'}$. Then, one of the following statements holds:
     \begin{enumerate}
      \item $a'_{s'}\leq b'_v$. Then, we can remove two pairs $(a'_{s'},b'_u)$ and $(b'_v,a_{s})$, and add $(a'_{s'},b'_v)$ and $(b'_u,a_{s})$. Obviously, we get a smaller cost (Fig. \ref{fig:j7}(a)). Contradiction.
       \item $b'_v\leq a'_{s'}\leq b'_u$. Then, two pairs $(a'_{s'},b'_u)$ and $(b'_v,a_{s})$ contradict the optimality (Fig. \ref{fig:j7}(b)).
       \item $b'_u\leq a'_{s'}\leq a_s$. This case contradicts the optimality of $C(a_{s},deg(a_{s}))$ (and $MM(a_{s},deg(a_{s}))$), since $deg(a_{s})\geq 2$ and $b'_v$ has been matched to $a_{s}$ while there existed a more near point $a'_{s'}$ with $deg(a'_{s'})<Cap(a'_{s'})$ (Fig. \ref{fig:j7}(c)).
     \end{enumerate}
 Notice that $b'_u=MA(a_s,deg(a_s))$, since when the point $a_s$ is examined (for computing $C(a_s,h)$ for $1\leq h\leq deg(a_s)$), we seek the points $p\leq a_s$ from the right to the left. Thus,

\begin{equation}
\nonumber
\begin{split}
C(b_1,1)=&b_1-a_s\\&+C(a_s,deg(a_s)-1).
\end{split}
\end{equation}

\begin{figure*}
\vspace{-0.5cm}
\hspace{0cm}
\includegraphics[width=1\textwidth]{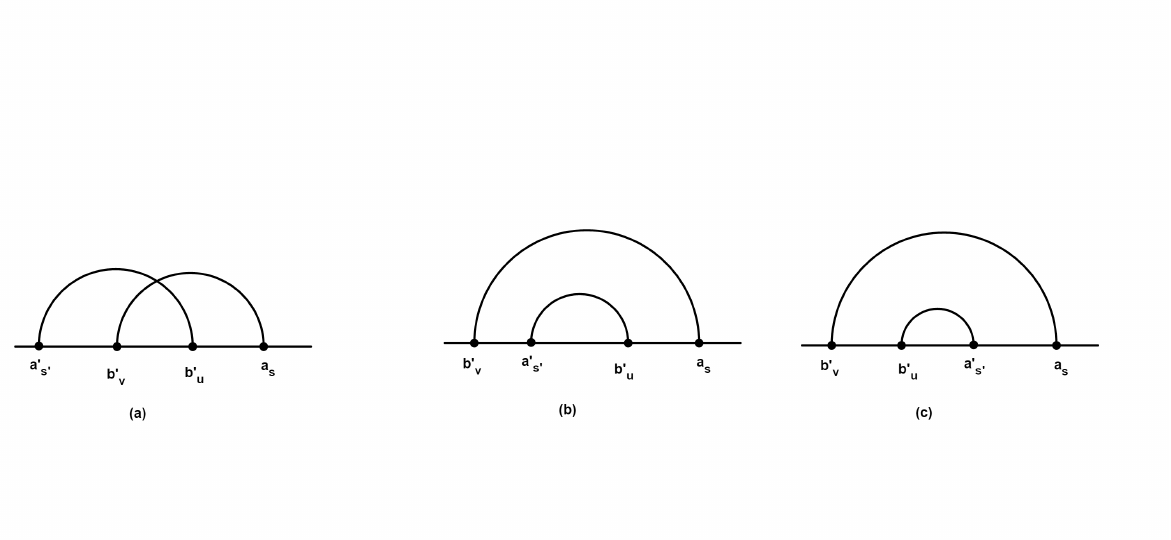}
\vspace{-1.5cm}
\caption{Suboptimal matchings. }
\label{fig:j7}       
\end{figure*}

  \item [Subcase 1.2:]$deg(a_s)<Cap(a_s)$. Then, one of the following subcases arises (Lines 5--16 of Algorithm \ref{Case3}):
  \begin{enumerate}
    \item [Subcase 1.2.1:]$deg(a_s)=1$. Then, we must match $b_1$ to $a_s$ and, moreover, examine that whether $a_s$ decreases the cost of the OLCMM or not, i.e. whether $C(b_1,1)$ includes $C(a_s,1)$ or not (Lines 6--13 of Algorithm \ref{Case3}). Thus,

  \begin{equation}
  \nonumber
  \begin{split}
  C(b_1,1)=&b_1-a_s+\min(C(a_s,1),\\
 &C(a_{s-1},deg(a_{s-1})).
  \end{split}
  \end{equation}

    \item [Subcase 1.2.2:]$deg(a_s)>1$. By Lemma \ref{lem3-3}, we simply match $b_1$ to $a_s$ (Lines 15--16 of Algorithm \ref{Case3}). Thus,
    $$C(b_1,1)=C(a_s,deg(a_s))+b_1-a_s.$$
  \end{enumerate}

\end{enumerate}

\item [Case 2:]If $i>1$, we use the following claim.

\begin{Claim}{}
Let $next1(w')\in A_{w'}$ be the largest point $p\in A_{w'}$ with $deg(p)<Cap(p)$ (if exists) and $next2(w')$ be the largest point of the following set:
$$M'=\{p\in A_{w'}:\{MA(p,h):MA(p,h)<p\}_{h=1}^{deg(p)}\neq \emptyset\},$$ (i.e., the largest point $p\in A_{w'}$ that has been matched to at least one smaller point $q\leq p$). Also, let $p'\in M'$ such that $p'<next2(w')$. Then, in a minimum-cost OLCMM, $b_i$ can be matched to either $next1(w')$ or $next2(w')$, but not to $p'$.
\label{claimnew1}
\end{Claim}

\begin{proof}
The claim can be easily proved by Observation \ref{observationnew2}.
\end{proof}

Assume that initially $next1(w)=next2(w)=a_s$ for all $w\geq 0$ (Line 10 of Algorithm \ref{OLCMM}). Now, we search $A_{w'}$ until one of the following cases arises (Algorithm \ref{FindNext}):
\begin{itemize}
  \item We reach the point $next1(w')$ with $deg(next1(w'))<Cap(next1(w'))$ (Lines 7--11 of Algorithm \ref{FindNext}). Then, by Claim \ref{claimnew1}, we search $A_{w'}$ to find $next2(w')$ (Lines 9--10 of Algorithm \ref{FindNext}). Then, we let $j=0$ and exit from the while loop (Line 11 of Algorithm \ref{FindNext}).
  \item We reach the point $next2(w')$ that has been matched to at least one smaller point $q\leq next2(w')$ (Lines 12--13 of Algorithm \ref{FindNext}). In this case, we assume that $next1(w')=a'_{j-1}$. We also let $j=0$ to exit from the while loop.
\end{itemize}

Observe that the conditions $s'>0$ and $s''>0$ in Lines 3 and 15 of Algorithm \ref{Find}, respectively, means that we could find such a point $next1(w')$ or $next2(w')$, since in Line 1 of Algorithm \ref{FindNext}, we set $s'=0$ and $s''=0$. Note that as Step 1, this case does not contradict the optimality of $C(b_{i-1},deg(b_{i-1}))$ (and $MM(b_{i-1},deg(b_{i-1}))$). Now, two subcases arise:

\begin{enumerate}

\item [Subcase 2.1.]We reach a point $a'_{s'}$ such that $deg(a'_{s'})<Cap(a'_{s'})$ (Lines 3--14 of Algorithm \ref{Find}). Then, we distinguish between two subcases:

\begin{enumerate}

\item [Subcase 2.1.1.]$1<deg(a'_{s'})<Cap(a'_{s'})$. Since $deg(a'_{s'})>1$, by Lemma \ref{lem3-3}, $b_i$ is matched to $a'_{s'}$ (Lines 4--6 of Algorithm \ref{Find}):

$$C(b_i,1)=C(b_{i-1},1)+b_i-a'_{s'}.$$

\item [Subcase 2.1.2.]$deg(a'_{s'})=1$. Let $next2(w')=a'_{s''}$, i.e. $a'_{s''}$ is the largest point in $A_{w'}$ that has been matched to at least one smaller point. Then, one of the following statements holds:

  \begin{itemize}
  \item There does not exist such a point $a'_{s''}$, i.e. $s''=0$ (Lines 4--6 of Algorithm \ref{Find}). Then, by Claim \ref{claimnew1}, we have:
        $$C(b_i,1)=C(b_{i-1},1)+b_i-a'_{s'}.$$
  \item There exists such a point $a'_{s''}$. Then, either $deg(a'_{s''})>1$ or $deg(a'_{s''})=1$. In the first case, from Lemma \ref{lem3-3} and Claim \ref{claimnew1}, we match $b_i$ to $a'_{s'}$ (Lines 4--6 of Algorithm \ref{Find}). And in the second one, by Claim \ref{claimnew1}, we check that whether $b_i$ must be matched to $a'_{s'}$ or $a'_{s''}$, i.e., we must verify that whether $C(b_i,1)$ includes $C(a'_{s''},1)$ or not (Lines 7--14 of Algorithm \ref{Find}). Thus, we have:

\begin{equation}
\nonumber
\begin{split}
C(b_i,1)=&C(b_{i-1},1)
   +\min(b_i-a'_{s'},\\&b_i-a'_{s''}-C(a'_{s''},1)\\&+C(a'_{s''},0)).
\end{split}
\end{equation}

 \item [Subcase 2.2.]We reach the point $a'_{s''}$ that has been matched to at least one smaller point $p\leq a'_{s''}$ (Lines 15--19 of Algorithm \ref{Find}), which implies that $deg(a'_{s''})=Cap(a'_{s''})$. Let $b'_u \leq a'_{s''}$ be the smallest point that has matched to $a'_{s''}$. And, let $v$ be the number of the points $p \leq a'_{s''}$ that have been matched to $a'_{s''}$. Notice that $b'_u=MA(a'_{s''},v)$, since when we want to compute $C(a'_{s''},h)$ for the point $a'_{s''}$ for $1\leq h\leq deg(a'_{s''})$, we seek the points of the previous partitions from the right to the left. In this subcase, as Subcase 1.1, by Observation \ref{observationnew2}, we must remove the pair $(b'_u,a'_{s''})$ and add the pair $(a'_{s''},b_i)$.



 Thus,
\begin{equation}
\nonumber
\begin{split}
C(b_i,1)=&C(b_{i-1},1)+b_i-a'_{s''}\\&-C(a'_{s''},v)
+C(a'_{s''},v-1).
\end{split}
\end{equation}




      \end{itemize}
\end{enumerate}

\end{enumerate}

\end{enumerate}
\end{itemize}

Now, if $deg(b_i)<1$ which means that we could not find any point in $A_{w'}$ for $b_i$ to be matched to, we must continue the search. Thus, we let $w'=w'-1$. Then, if $w'\geq 1$, we let $w'=ns(w')$ and repeat the while loop, otherwise we let $w'=-1$ to exit from the while loop (Lines 26--28 of Algorithm \ref{OLCMM}).

 \end{itemize}

Observe that it is possible that after executing Steps 1--2, we still have $deg(b_i)=0$. This means that we could not match $b_i$ to any point $p\in S\cup T$ with $p\leq b_i$, which implies that there does not exist any OLCMM for the points $\{p\in S\cup T:p\leq b_i\}$. Moreover, recall that $C(p,k)=null$ for $p\in S\cup T$ and $1\leq k\leq Cap(p)$ implies that $C(p,k)=\infty$. Thus, the conditions below are used to ensure that our algorithm can determine the case where there does not exist any OLCMM for the points $\{p\in S\cup T:p\leq b_i\}$:

\begin{itemize}
  \item $C(a_s,deg(a_s)-1)\neq null$ (Line 2 of Algorithm \ref{Case3}).
  \item $C(a_{s-1},deg(a_{s-1}))\neq null$ (Line 6 of Algorithm \ref{Case3}).
  \item $C(a'_{s''},0)=null$ (Line 4 of Algorithm \ref{Find}).
  \item $C(a'_{s''},v-1)\neq null$ (Line 17 of Algorithm \ref{Find}).
\end{itemize}

Observe that we compute $C(p,k)$ for $k \in\{1,2,\dots,Cap(p)\}$ only if the $k$th capacity of the point $p\in S\cup T$ decreases the cost of the OLCMM corresponding to $\{q\in S\cup T:q\leq p\}$; otherwise we only compute $C(p,1)$. Recall that in Step 1, each point $q\in S\cup T$ might be examined only once by one of the points $p\in S\cup T$ with $q<p$. Thus, the number of the elements $C(p,k)$ in the linked list is not more than $2*n$ in the worst case.
 \end{proof}




\algsetblock[Name]{Initial}{}{3}{1cm}
\alglanguage{pseudocode}
\begin{algorithm2e}

\caption{Step1($M$, $MM$, $S$, $T$, $i$, $A_0,A_1, \dots$)}
\label{Case1.1}


\If{$i=1$}{
    $j=s$\;

   \Call{Proc}{$b_1$,$A_{w}$,$j$,$w+1$}\;

}

$w''=w+1$ and $t'=i-1$\;

\While{$(w''\geq 1)\land(deg(b_i)<Cap(b_i))\land (continue(w'')=True)$}{

  Let $A_{w'}$ be the partition containing $MA(b_{t'},deg(b_{t'}))$\;
  Let $MA(b_{t'},deg(b_{t'})$ be the $s'$th point of $A_{w'}=\{a'_1,a'_2,\dots\}$\;
   $j'=s'-1$\;
    \Call{Proc}{$b_i$,$A_{w'}$,$j'$,$w''$}\;

  \lIf{$continue(w'')=True$}{
  $w''=w'-1$ and $t'=\vert A_{w''}\vert$}


  }

   \If{$(w''< 1)\lor (continue(w'')=False)$}{
   $continue(w+1)=False$\;
   }

\end{algorithm2e}

\algsetblock[Name]{Initial}{}{3}{1cm}
\alglanguage{pseudocode}
\begin{algorithm2e}

\caption{Proc($b_i$,$A_{w'}$,$j'$,$w''$)}
\label{Proc1}

   \While {$(j'\geq 1)\land (deg(b_i)<Cap(b_i))$}{
     \If{$deg(a'_{j'})\leq1$}{
      \If{$deg(a'_{j'})=0$}{
           Add the pair $(b_i,a'_{j'})$ to $M$\;
           $MM(b_i,deg(b_i))=M$\;
           $C(b_i,deg(b_i))=C(b_i,deg(b_i)-1)+b_i-a'_{j'}$\;}
      \ElseIf{$MA(a'_{j'},1)\leq a'_{j'}$}{
         \If{$C(a'_{j'},1)>C(a'_{j'-1},deg(a'_{j'-1}))+b_i-a'_{j'}$ }{
              Add the pair $(b_i,a'_{j'})$ to $M$ and remove $(a'_{j'},MA(a'_{j'},1))$\;
              $MM(b_i,deg(b_i))=M$\;
              $C(b_i,deg(b_i))=C(b_i,deg(b_i)-1)+b_i-a'_{j'}-C(a'_{j'},1)+C(a'_{j'-1},deg(a'_{j'-1}))$\;
              $j'=j'-1$\;}
         \lElse{
               $j'=0$, and $continute(w'')=False$}


             }

       \lElse{
           $j'=j'-1$}

            }

            \lElse{
            $j'=0$, and $continue(w'')=False$
            }


}

\end{algorithm2e}

\algsetblock[Name]{Initial}{}{3}{1cm}
\alglanguage{pseudocode}
\begin{algorithm2e}
\caption{Step2.1($M$, $MM$, $S$, $T$, $i$, $A_0,A_1, \dots$,$w'$)}
\label{Case2}


 $\{b'_1,b'_2,\dots,b'_{t'}\}=\{p\in A_{w'}:p< b_i\}$, and $v=deg(b'_{id(w')})$\;
\If {$id(w')>0$}{

       Let $\{m_1,m_2,\dots,m_v\}=\{MA(b'_{id(w')},1),MA(b'_{id(w')},2),\dots,MA(b'_{id(w')},v)\}$\;
       Add the pair $(b_i,m_v)$ to $M$ and remove $(b'_{id(w')},m_v)$ from $M$, $MM(b_i,1)=M$\;
      $C(b_i,1)=C(b_{i-1},deg(b_{i-1}))+b_i-b'_{id(w')}$\;
    \If{$deg(b'_{id(w')})=1$}{
          $id(w')=id(w')-1$\;}
        }
    \ElseIf{$1<deg(MA(b'_{t'},1))<Cap(MA(b'_{t'},1))$}{

        $m'=MA(b'_{t'},1)$\;
        Add the pair $(b_i,m')$ to $M$\;
        $MM(b_i,1)=M$\;
       $C(b_i,1)=C(b_{i-1},1)+b_i-m'$\;
       }

    \ElseIf{$i>1$}{
     \If{$1<deg(MA(b_{i-1},1))<Cap(MA(b_{i-1},1))$}{

        $m'=MA(b_{i-1},1)$\;
        Add the pair $(b_i,m')$ to $M$\;
        $MM(b_i,1)=M$\;
       $C(b_i,1)=C(b_{i-1},1)+b_i-m'$\;
       }
       }

 \end{algorithm2e}

\algsetblock[Name]{Initial}{}{3}{1cm}
\alglanguage{pseudocode}

\begin{algorithm2e}
\caption{Step2.2($M$, $MM$, $S$, $T$, $i$, $A_0,A_1, \dots$,$w'$)}
\label{Case3}
\SetAlgoVlined
     \If{$i=1$}{
           \If{$(deg(a_s)=Cap(a_s))\land (C(a_s,deg(a_s)-1)\neq null)$}{
                $C(b_1,1)=b_1-a_s+C(a_s,deg(a_s)-1)$\;
                 $M=MM(a_s,deg(a_s)-1)\cup (a_s,b_1)$, $MM(b_1,1)=M$\;

                 }
            \ElseIf{$deg(a_s)<Cap(a_s)$}{
                \If{($deg(a_s)=1)\land (C(a_{s-1},deg(a_{s-1}))\neq null)$}{

                    \If{$C(a_s,1)\leq C(a_{s-1},deg(a_{s-1}))$}
                    {
                    $C(b_1,1)=b_1-a_s+C(a_s,1)$\;
                     $M=M\cup (a_s,b_1)$\;
                    }

                    \Else
                    {
                    $C(b_1,1)=b_1-a_s+C(a_{s-1},deg(a_{s-1}))$\;
                     $M=MM(a_{s-1},deg(a_{s-1}))\cup (a_s,b_1)$\;
                    }

                     $MM(b_1,1)=M$\;

                     }
                \Else{
                     $C(b_1,1)=C(a_s,deg(a_s))+b_1-a_s$\;
                     $M=M\cup (a_s,b_1)$, $MM(b_1,1)=M$\;}

}
}

        \lElse{
             \Call{Find}{$M$, $MM$, $S$, $T$, $i$, $A_0,A_1, \dots$,$w'$}
          }

\end{algorithm2e}

\algsetblock[Name]{Initial}{}{3}{1cm}
\alglanguage{pseudocode}
\begin{algorithm2e}
\caption{Find($M$, $MM$, $S$, $T$, $i$, $A_0,A_1, \dots$,$w'$)}
\label{Find}

\SetAlgoVlined

          $s'=s''=0$\;

         ($s'$,$s''$)=\Call{FindNext}{$next1(w')$,$next2(w')$,$A_{w'}$}\;

         \If{$s'>0$}{
            \If{$(deg(a'_{s'})>1)\lor (s'' =0)\lor (deg(a'_{s''})>1)\lor (C(a'_{s''},0)=null)$ }{
                  $C(b_i,1)=C(b_{i-1},1)+b_i-a'_{s'}$\;
                  $M=M\cup (b_i,a'_{s'})$, $MM(b_i,1)=M$\;}
             \Else{

             \If{$b_i-a'_{s'}\leq b_i-a'_{s''}-C(a'_{s''},1)+C(a'_{s''},0)$}
             {
             $C(b_i,1)=C(b_{i-1},1)+b_i-a'_{s'}$\;
             $M=M\cup (b_i,a'_{s'})$\;}
             \Else{

             $C(b_i,1)=C(b_{i-1},1)+b_i-a'_{s''}-C(a'_{s''},1)+C(a'_{s''},0)$\;
             $M=M\cup (a'_{s''},b_i)$, $M=(M\setminus MM(a'_{s''},1))\cup MM(a'_{s''},0)$ \;


              }%

              $MM(b_i,1)=M$\;
}
}
         \ElseIf {$s''>0$}{
            Let $v$ be the number of the points of $\{MA(a'_{s''},h):MA(a'_{s''},h)\leq a'_{s''}\}_{h=1}^{deg(a'_{s''})}$\;
         \If{$C(a'_{s''},v-1)\neq null$}{

           $C(b_i,1)=C(b_{i-1},1)+b_i-a'_{s''}-C(a'_{s''},v)+C(a'_{s''},v-1)$\;
           $M=M\cup (a'_{s''},b_i)$, $M=(M\setminus MM(a'_{s''},v))\cup MM(a'_{s''},v-1)$, $MM(b_i,1)=M$\;
          }%

         }%


\end{algorithm2e}

\algsetblock[Name]{Initial}{}{3}{1cm}
\alglanguage{pseudocode}
\begin{algorithm2e}

\caption{FindNext($next1(w')$,$next2(w')$,$A_{w'}$)}
\label{FindNext}
\SetAlgoVlined
            $s'=s''=0$\;
            $A_{w'}=\{a'_1,a'_2,\dots\}$, and $a'_0=\max(p\in A_{w'-1})$\;
             Let $j'$ be the number of the points of $\{p\in A_{w'}:p\leq next1(w')\}$\;
             Let $j''$ be the number of the points of $\{p\in A_{w'}:p\leq next2(w')\}$\;
             $j=\max(j',j'')$\;

           \While {$j\geq 1$}{
              \If {$deg(a'_j)<Cap(a'_j)$}{
                    $next1(w')=a'_j$, $s'=j$, and $j=\min(j,j'')$\;
                  \While{$(j\geq 1)\land(\{MA(a'_{j},h):MA(a'_{j},h)\leq a'_{j}\}_{h=1}^{deg(a'_{j})}= \emptyset)$}
                     {
                       $j=j-1$\;
                    }
                    $next2(w')=a'_{j}$, $s''=j$, and $j=0$\;
                    }

              \ElseIf{$\exists p \in \{MA(a'_j,h)\}_{h=1}^{deg(a'_j)}$ s.t. $p\leq a'_j$}{
                    $next2(w')=a'_j$, $next1(w')=a'_{j-1}$, $s''=j$, and $j=0$\;}
              \lElse{
           $j=j-1$}
         }

      \Return($s'$,$s''$)\;

\end{algorithm2e}

\FloatBarrier

\end{document}